\documentclass[aps,prb,twocolumn,floatfix,superscriptaddress]{revtex4-2}
\usepackage[T1]{fontenc}
\usepackage{amsmath, amssymb, amsfonts, mathrsfs, bm}
\usepackage{graphicx}
\usepackage{physics}
\usepackage{times}
\usepackage{empheq}
\usepackage{bbold}
\usepackage{dsfont}
\usepackage{xcolor}
\usepackage{cancel}
\usepackage{standalone}
\usepackage{stmaryrd}
\usepackage{microtype}
\usepackage{comment}
\usepackage{tikz}
\usetikzlibrary{decorations.markings, shapes, positioning, arrows}
\usepackage{slashed}
\usepackage{siunitx}
\usepackage[colorlinks=true, allcolors=black, citecolor=blue, linkcolor=blue, urlcolor=blue]{hyperref}
\usepackage[capitalize]{cleveref}
\usepackage{xspace}

\renewcommand{\vec}[1]{\boldsymbol{#1}}

\newcommand{\br}{\vec{r}}
\newcommand{\bk}{\vec{k}}
\newcommand{\ie}{\emph{i.e.}\xspace}

\newcommand{\kp}{$\vec{k}\cdot\vec{p}$\ }
\newcommand{\kpns}{$\vec{k}\cdot\vec{p}$}
\newcommand\zero{\mathbb{0}}
\newcommand\one{\mathbb{1}}

\begin{document}

\title{Valley physics in the two bands k.p model for SiGe heterostructures and spin qubits}

\author{Tancredi Salamone}
\affiliation{Univ. Grenoble Alpes, CEA, Leti, F-38000, Grenoble, France.}
\author{Biel Martinez Diaz}
\affiliation{Univ. Grenoble Alpes, CEA, Leti, F-38000, Grenoble, France.}
\author{Jing Li}
\affiliation{Univ. Grenoble Alpes, CEA, Leti, F-38000, Grenoble, France.}
\author{Lukas Cvitkovich}
\affiliation{Institute of Theoretical Physics, University of Regensburg, 93040 Regensburg, Germany.}
\author{Yann-Michel Niquet}
\email{yniquet@cea.fr}
\affiliation{Univ. Grenoble Alpes, CEA, IRIG-MEM-L\_Sim, Grenoble, France.}%

\date{\today}

\begin{abstract}
We discuss the choice and implementation of inter-valley potentials in the so-called two bands \kp model for the opposite $X$, $Y$ or $Z$ valleys of silicon. We focus on the description of valley splittings in Si/SiGe heterostructures for spin qubits, with a particular attention to alloy disorder. We demonstrate that the two bands \kp model reproduces the valley splittings of atomistic tight-binding calculations in relevant heterostructures (SiGe spikes, wiggle wells...), yet at a much lower cost. We show that the model also captures the effects of valley-orbit mixing and yields the correct inter-valley dipole matrix elements that characterize manipulation, dephasing and relaxation in spin/valley qubits. We simulate a realistic Si/SiGe spin qubit device as an illustration, and discuss electron-phonon interactions in the two bands \kp model. Beyond spin qubits, this model enables efficient simulations of SiGe heterostructure devices where spin and valley physics are relevant.
\end{abstract}

\maketitle

\section{Introduction}

Spin qubits in semiconductor quantum dots have established themselves as a compelling platform for quantum computing and simulation \cite{Burkard2023, Vandersypen2017,Takeda2022,Hsiao2024,Fariña2025}. Coherent and high fidelity single and two-qubit gates have been demonstrated with various carriers and materials, such as electrons and holes in silicon metal-oxide-semiconductor (MOS) devices \cite{Maurand2016,Camenzind2022,Klemt2023,Geyer2024,Steinacker2025}, electrons in Si/SiGe heterostructures \cite{Philips2022,Noiri2022,Mills2022,Xue2022,Weinstein2023,Unseld2023,George2025,Wu2025}, and holes in Ge/GeSi heterostructures \cite{Hendrickx2021,John2024,Wang2024,Zhang2025}. These heterostructures are at the forefront of the most advanced realizations owing to the high quality of epitaxial interfaces \cite{Sammak2019,Lawrie2020,Martinez2022,Martinez2024,Martinez2025}.

One of the main challenges faced by electron spin qubits in Si/SiGe heterostructures is the management of the valley splittings. The conduction band of bulk silicon is, indeed, highly degenerate, with six equivalent valleys near the $\pm X$, $\pm Y$ and $\pm Z$ points of the first Brillouin zone \cite{Zwanenburg2013}. Although the degeneracy between $X$, $Y$ and $Z$ valleys can easily be lifted by strains and confinement, the splitting $\Delta$ between the opposite ground-state valleys (e.g., $\pm Z$) remains usually small (tens of micro-electronvolts) \cite{Borselli2011,Yang2013,Gamble2016,Scarlino2017,Mi2017,Neyens2018,Ferdous2018,Hollmann2020,Chen2021,Spence2022} and highly variable from dot to dot \cite{DegliEsposti2024,Volmer2024,Peña2024}. This provides a leakage channel for the spin qubits that degrades the coherence and fidelities \cite{Langrock2023,Losert2024,Volmer2025}, especially when $\Delta$ is comparable if not even smaller than the Zeeman splitting. 

The quasi-degeneracy between the opposite $Z$ valleys can actually be lifted by rapidly varying potentials with significant Fourier components at wave numbers $q_z=\pm 2k_0$, where $\pm k_0$ are the positions of the $\pm Z$ valleys along the reciprocal $z$ axis \cite{Friesen2005,Friesen2006,Goswami2007,Friesen2007,Saraiva2009,Saraiva2011}. This ``$2k_0$'' theory sparked several proposals to enhance valley splittings, such as the introduction of a GeSi spike in the Si well \cite{McJunkin2021}, or the so-called ``wiggle'' wells \cite{Gradwohl2025,Feng2022,McJunkin2022} with an oscillating Ge fraction. In these designs, the valley splittings can reach a few hundreds of micro-electronvolts, efficiently limiting the admixture of spin and valley degrees of freedom. Earlier studies also underlined the role of alloy disorder \cite{PaqueletWuetz2022,Losert2023,Lima2023}, whose $q_z=\pm 2k_0$ components compete with those of the intended Ge concentration profile. 

The most accurate electronic structure methods for valley splittings in SiGe heterostructures are atomistic descriptions, such as density functional theory \cite{Cvitkovich2025} and tight-binding models \cite{Boykin2004,Nestoklon2006,Klimeck2007_1,Klimeck2007_2,Chutia2008,Abadillo2018}, which can capture details such as alloy disorder \cite{PaqueletWuetz2022,Losert2023}. However, these methods are numerically expensive, and are thus hardly suitable for the modeling of realistic device-scale structures that may contain tens of millions of atoms. Continuous medium approximations, such as the effective mass equation, are much more appropriate for that purpose.

The original effective mass equations do, however, decouple the opposite valleys, hence give rise to zero valley splittings. The couplings between the opposite valleys may nevertheless be treated as a perturbation of the effective mass solution (the $2k_0$ theory) \cite{Nestoklon2006,Friesen2007,Saraiva2009,Culcer2010,Saraiva2011}. Recent works have thus introduced model inter-valley potentials that capture the effects of alloy disorder within this framework \cite{Lima2023,Lima2024}. There has been, moreover, various proposals for refinements and extensions of the $2k_0$ theory \cite{Thayil2025,Binder2025}. It would generally be preferable to deal with the inter-valley coupling non-perturbatively in order to catch, in particular, valley-orbit mixing (the fact that the valley wave functions may have different envelopes) \cite{Friesen2010,Gamble2013}. This is essential to obtain, e.g., the correct inter-valley dipole matrix elements. 

The two bands \kp model for the conduction bands of Si and Ge is an extension of the effective mass equations that explicitly couples pairs of opposite valleys \cite{Hensel1965,Sverdlov2007,Sverdlov2008,Osintsev2014,Osintsev2015,Sverdlov2015}. This model accounts for various physics beyond the effective mass approximation, such as the effects of shear strains on the valleys, or the bulk Dresselhaus spin-orbit interactions. It originally features a kinetic valley-orbit mixing term, but does not describe valley splitting by the potential. In this work, we further extend the two bands \kp model by including such an inter-valley potential. We rewrite for that purpose the $2k_0$ theory in the frame of the two bands \kp model. We discuss the choice of the inter-valley potential in SiGe alloys, and its implementation in finite-difference solvers \cite{Mattiussi2000}. We then validate the model against atomistic tight-binding calculations in various Si/SiGe heterostructures of interest (SiGe spikes, wiggle wells, ...). We show that the two bands \kp model reproduces the TB valley splittings and inter-valley dipole matrix elements in the presence of alloy disorder. Finally, we illustrate the relevance of this model with a simulation of a realistic qubit device. We discuss on this occasion charge and spin-phonon interactions in the two bands \kp model.

We introduce the two bands \kp model and discuss the choice and implementation of an inter-valley potential in \cref{sec:methodology}, then validate the model against TB calculations in \cref{sec:validation}, and explore valley-orbit mixing and dipole matrix elements in \cref{sec:voc_dip}. Finally, we apply this modeling framework to realistic qubit devices in \cref{sec:qubits}.

\section{Methodology}
\label{sec:methodology}

In this section, we review the valley splitting in the $2k_0$ theory, then introduce the two bands \kp model (with and without spin), and discuss the implementation of valley-orbit mixing potentials in this model. We then consider the particular case of SiGe heterostructures, and the treatment of alloy disorder. 

\subsection{The valley splitting in the $2k_0$ theory}

We consider an electron moving in silicon in a potential $V(\br)$ slowly varying on the scale of the diamond unit cell. In the simplest effective mass approximation, the eigenwave functions $\psi_{\pm}(\br)$ of the $\pm Z$ valleys read
\begin{equation}
    \psi_{\pm}(\br)=e^{\pm ik_0z}\varphi(\br)u_\pm(\br)\,,
    \label{eq:psi_vs}
\end{equation}
where $u_\pm(\br)$ are the Bloch functions of the two valleys at wave vectors $\bk_\pm=\pm k_0\hat{\vec{z}}$ (with $k_0=0.85\times 2\pi/a$, $a$ the lattice parameter, and $\hat{\vec{z}}$ the unit vector along $z=[001]$), and $\varphi$ is an envelope function. The latter is an eigensolution of the anisotropic effective mass Hamiltonian
\begin{equation}
    H=-\frac{\hbar^2}{2}\left(\frac{1}{m_t^*}\frac{\partial^2}{\partial x^2}+\frac{1}{m_t^*}\frac{\partial^2}{\partial y^2}+\frac{1}{m_l^*}\frac{\partial^2}{\partial z^2}\right)+V(\br)\,,
    \label{eq:ema}
\end{equation}
with $m_l^*=0.916\,m_0$ the longitudinal effective mass along $z$, $m_t^*=0.191\,m_0$ the transverse effective mass along $x$ and $y$, and $m_0$ the free electron mass. The equations for the $X$ and $Y$ valleys can be obtained by a permutation of the $x$, $y$ and $z$ axes. Each eigenstate $|\psi_{\pm}\rangle$ is thus twice valley-degenerate. 

This degeneracy is lifted if the potential $V(\br)$ can couple opposite valley wave functions. To lowest order in perturbation, the valley splitting indeed reads $\Delta=2|J|$, with \cite{Saraiva2009,Saraiva2011}:
\begin{equation}
    J=\langle\psi_+|V|\psi_-\rangle\,.
    \label{eq:Jinter}
\end{equation}
To separate the contributions from the envelope and periodic Bloch functions, we can conveniently expand the latter in Fourier series \cite{Saraiva2009}
\begin{equation}
    u_\pm(\br)=\sum_{\vec{G}}c_{\pm}(\vec{G})e^{i\vec{G}\cdot\br}\,,
    \label{eq:blochs_exp}
\end{equation}
where $\vec{G}$ is a vector of the reciprocal lattice. Note that in the absence of spin-orbit coupling, time-reversal symmetry imposes $u_-(\br)=u_+^*(\br)$, thus $c_-(\vec{G})=c_+^*(-\vec{G})$. As $V(\br)$ and $\varphi(\br)$ are assumed to be slowly varying over the unit cell, we can approximate the integrals over in-plane coordinates in \cref{eq:Jinter} as:
\begin{align}
    \int dxdy&\,e^{i(\vec{G}-\vec{G}')\cdot\br}V(\br)\left|\varphi(\br)\right|^2\simeq\delta_{G_x,G_x'}\delta_{G_y,G_y'}\times \nonumber \\
    &e^{i(G_z-G_z')z}\int dxdy\,V(\br)\left|\varphi(\br)\right|^2\,.
\end{align}
We thus reach
\begin{equation}
    J=\sum_{\vec{G},\vec{G}'}c_+^*(\vec{G}')c_-(\vec{G})\delta_{G_x,G_x'}\delta_{G_y,G_y'}I(G_z-G_z'-2k_0)\,, \label{eq:Jfull}
\end{equation}
where:
\begin{equation}
    I(k_z)=\int d^3\br\,e^{ik_z z}V(\br)\left|\varphi(\br)\right|^2\,.
    \label{eq:IGzGzp}
\end{equation}
The integral $|I(k_z)|$ is also expected to decrease with large, increasing $|k_z|$, so that the sum in \cref{eq:Jfull} shall be dominated by the smallest $|G_z-G_z'-2k_0|$, which are $G_z'=G_z$ and $G_z-G_z'=4\pi/a$. The latter does not, however, contribute for symmetry reasons \cite{note4pia}. We are hence left with
\begin{equation}
J\approx A\int d^3\br\,e^{-2ik_0 z}\,V(\br)\left|\varphi(\br)\right|^2\,, 
\label{eq:Jfinal}
\end{equation}
where
\begin{equation}
A=\sum_{\vec{G}} c^*_+(\vec{G})\, c_-(\vec{G})=\frac{1}{\Omega_0}\langle u_+|u_-\rangle_{\Omega_0}
\label{eq:Acoeff}
\end{equation}
and $\langle\cdot|\cdot\rangle_{\Omega_0}$ is a scalar product over the unit cell with volume $\Omega_0$. Using {\it ab initio} Bloch functions \cite{Saraiva2011} yields $A=-0.26$ \footnote{Note that the phase of $A$ depends on the choice of phase for the Bloch functions. This (global) phase is however irrelevant as long as all inter-valley matrix elements are computed consistently.}. In this ``$2k_0$'' theory, the valley splitting is thus proportional to the Fourier transform of the density-weighted potential at the inter-valley wave vector $\vec{q}=2k_0\hat{\vec{z}}$ ($2k_0\approx 20$\,nm$^{-1}$) \footnote{Note that \cref{eq:Jfinal} is not strictly invariant by a rigid shift of the potential $V(\br)$ (since the squared envelope function may also have small $2k_0$ components), although shifts of the order of the 100 meV only change the valley splitting by 2-3\% [as tested with the potential of \cref{eq:modelpot}]. This results from the fact that the valley wave functions defined by \cref{eq:psi_vs} are not strictly orthogonal.}. The potential $V(\br)$ must, therefore, have sizable Fourier amplitudes at that wave vector to achieve large valley splittings. This is the rationale for proposals, such as wiggle wells \cite{Feng2022,McJunkin2022}, where the heterostructure is carefully designed to enhance the $2k_0$ component of the vertical confinement potential.

\subsection{The two bands k.p model}
\label{sec:2kp}

We can introduce the functions
\begin{subequations}
\begin{align}
    \tilde{u}_-(\br)&=u_+(\br) \\
    \tilde{u}_+(\br)&=e^{-i\tfrac{4\pi}{a}z}u_-(\br)
    \label{eq:utilde}
\end{align}
\end{subequations}
and refactor $\psi_\pm$ as 
\begin{equation}
    \psi_\pm(\br)=e^{\pm i\tfrac{2\pi}{a}z}e^{\mp ik_0'z}\varphi(\br)u_\pm(\br)=e^{i\tfrac{2\pi}{a}z}\tilde{\psi}_\mp(\br) \label{eq:wfn2kp}
\end{equation}
with $k_0'=2\pi/a-k_0=0.15\times 2\pi/a$ and
\begin{equation}
    \tilde{\psi}_\pm(\br)=e^{\pm ik_0'z}\varphi(\br)\tilde{u}_\pm(\br)=\tilde{\varphi}_\pm(\br)\tilde{u}_\pm(\br), \label{eq:tildewf}
\end{equation}
where we have defined new envelope functions $\tilde{\varphi}_\pm(\br)=e^{\pm ik_0'z}\varphi(\br)$ that lump the (rather slowly varying) phase factors $e^{\pm ik_0'z}$. We can thus rewrite \cref{eq:Jfinal} as:
\begin{equation}
    J=A\int d^3\br\,e^{-i\frac{4\pi}{a}z}\tilde{\varphi}_-^*(\br)V(\br)\tilde{\varphi}_+(\br)\,. \label{eq:Jtilde}
\end{equation}
Note that the phase factor $e^{-i\frac{4\pi}{a}z}$ in this integral changes sign every $\delta z=a/4$, \ie every Si monolayer.

\begin{figure}
    \centering
    \includegraphics[width=\columnwidth]{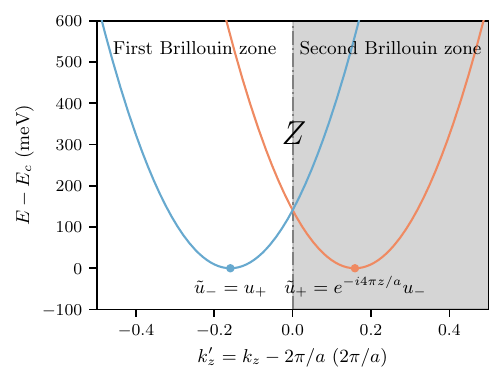}
    \caption{The conduction band structure of silicon in an extended zone scheme around $Z$. The blue parabola is the $+Z$ valley centered around $k_z=0.85\times 2\pi/a$. The orange parabola is the replica of the $-Z$ valley centered around $k_z=-0.85\times 2\pi/a$ in the second Brillouin zone. Both bands are degenerate at $Z$. The two bands \kp model expands the wave functions on the degenerate Bloch functions $\hat{u}_\pm(\br)\equiv\tilde{u}_\pm(\br)$ at $Z$. $E_c$ is the conduction band edge energy.}
    \label{fig:bs}
\end{figure}

We emphasize that $\tilde{u}_+$ is still periodic over the diamond lattice and is in fact the Bloch function of the conduction band at wave vector $\bk=(2\pi/a+k_0')\hat{\vec{z}}$ in the second Brillouin zone (while $\tilde{u}_-$ is the Bloch function at $\bk=(2\pi/a-k_0')\hat{\vec{z}}$, see \cref{fig:bs}). \cref{eq:wfn2kp} and \cref{eq:tildewf} are in fact very similar to the expression of the wave functions in the so-called two-bands \kp model for the $Z$ valleys of silicon \cite{Hensel1965,Sverdlov2007,Sverdlov2008}. This model indeed expands the conduction band wave functions as
\begin{equation}
\psi(\br)=e^{i\tfrac{2\pi}{a}z}\left[\hat{\varphi}_-(\br)\hat{u}_-(\br)+\hat{\varphi}_+(\br)\hat{u}_+(\br)\right]
\end{equation}
where $\hat{u}_\pm(\br)$ are the degenerate Bloch functions at $Z$ (that may \emph{a priori} differ from the Bloch functions at $\bk=(2\pi/a\pm k_0')\hat{\vec{z}}$). The envelopes $\hat{\Phi}=[\hat{\varphi}_-,\hat{\varphi}_+]$ of the eigenwave functions of the two-bands \kp model fulfill the system of differential equations
\begin{equation}
H_\mathrm{2kp}^\mathrm{kin}(-i\vec{\nabla})\hat{\Phi}+V(\br)I_2\hat{\Phi}=E\hat{\Phi}
\label{eq:h2kp}
\end{equation}
where $I_2$ is the $2\times2$ identity matrix and
\begin{equation}
H_\mathrm{2kp}^\mathrm{kin}(\bk')=
\begin{bmatrix}
H_\mathrm{EMA}(\bk'+k_0'\hat{\vec{z}}) & H_\mathrm{VOM}(\bk') \\
H_\mathrm{VOM}(\bk') & H_\mathrm{EMA}(\bk'-k_0'\hat{\vec{z}})
\end{bmatrix}
\label{eq:h2kpkin}
\end{equation}
with;
\begin{subequations}
\label{eq:EMAinter}
\begin{align}
H_\mathrm{EMA}(\bk')&=E_c+\frac{\hbar^2}{2}\left(\frac{{k'_x}^2}{m_t^*}+\frac{{k'_y}^2}{m_t^*}+\frac{{k'_z}^2}{m_l^*}\right) \\
H_\mathrm{VOM}(\bk')&=-\frac{\hbar^2}{M}k'_xk'_y\,.
\end{align}    
\end{subequations}
Here $E_c$ is the conduction band edge energy of the $Z$ valleys and $1/M=1/m_t^*-1$.

In bulk Si, the Bloch functions at $\bk=\bk'+2\pi/a\hat{\vec{z}}$ are the eigenvectors of $H_\mathrm{2kp}(\bk')$. Since $H_\mathrm{2kp}(\pm k'_0\hat{\vec{z}})$ is diagonal, the Bloch functions of the 2 bands \kp model and of \cref{eq:wfn2kp} can be formally identified (up to phase factors), so that $\hat{u}_\pm$ can be replaced by $\tilde{u}_\pm$ and $\hat{\varphi}_\pm$ by $\tilde{\varphi}_\pm$ \cite{note4pia}.

In \cref{eq:EMAinter}, $H_\mathrm{EMA}$ is nothing else than the effective mass Hamiltonian (shifted in reciprocal space to account for the $e^{\pm ik_0'z}$'s lumped into the $\tilde{\varphi}_\pm$'s), while $H_\mathrm{VOM}(\bk')$ is a $\emph{kinetic}$ valley-orbit mixing Hamiltonian. $H_\mathrm{VOM}(\bk')$ is responsible for significant band warping and non-parabolicity along the transverse mass axes (an asset of this model with respect to the effective mass equation) \cite{Sverdlov2008}. This kinetic term is, on the one hand, missing in the $2k_0$ theory; but the effects of the potential $V(\br)$ on valley mixing [\cref{eq:Jtilde}] are, on the other hand, missing in $H_\mathrm{2kp}$. We next discuss how to implement such a valley(-orbit) mixing potential in the two-bands \kp model.

\subsection{Implementation of an inter-valley potential in the two bands k.p model} 
\label{sec:implementation2kp}

\begin{figure}
    \centering
    \includegraphics[width=\columnwidth]{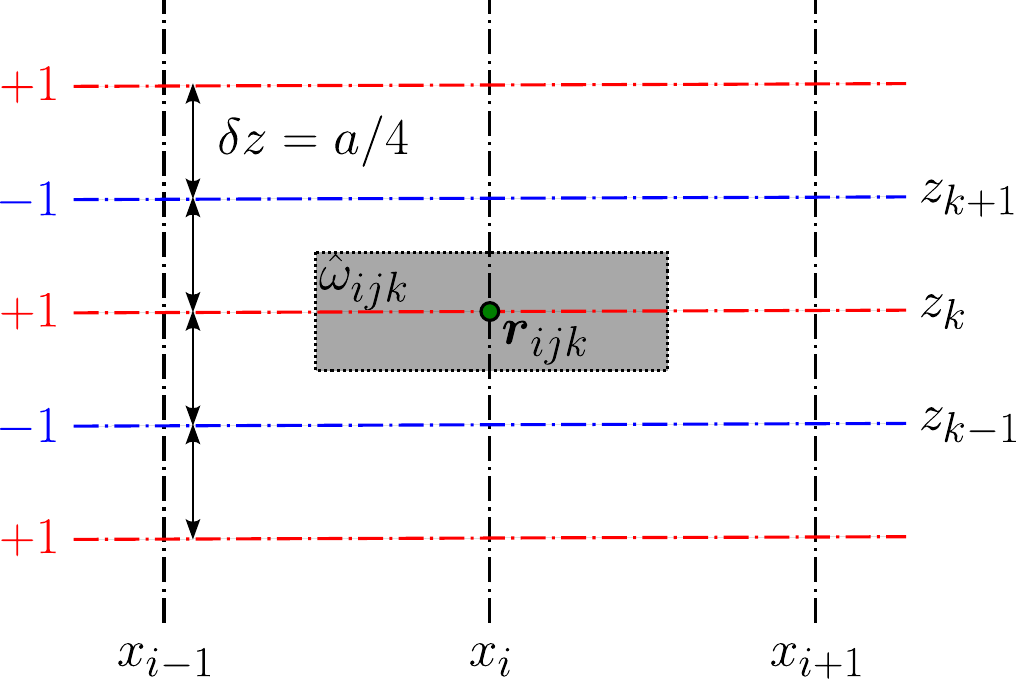}
    \caption{Cross-section (in the $xz$ plane) of an illustrative finite-differences mesh used to solve the two bands \kp equations. The spacing between successive mesh lines along $z$ is $\delta z=a/4$ (the bare distance between monolayers). The inter-valley potential $V_\mathrm{inter}$ is multiplied by $+1$ (red) or $-1$ (blue) on each monolayer.}
    \label{fig:mesh}
\end{figure}

\cref{eq:h2kp} can be solved using finite differences \cite{Mattiussi2000} on a rectilinear mesh $\br_{ijk}$ (the product of possibly non-homogeneous meshes $\{x_i\}$ along $x$, $\{y_j\}$ along $y$, and $\{z_k\}$ along $z$, see \cref{fig:mesh}). Neglecting $H_\mathrm{VOM}$ to begin with, the two valleys decouple and the valley splitting is given, to first order in perturbation, by \cref{eq:Jtilde}. We emphasize that \cref{eq:Jtilde} is equivalent to \cref{eq:Jfinal} and is thus the Fourier transform of $f(\br)\equiv V(\br)\left|\varphi(\br)\right|^2$ at wave number $q_z=2k_0$. Therefore, if $f(\br)$ has negligible Fourier components beyond $q_z=4\pi/a$ \footnote{We point out that this (Shannon-Nyquist) constraint on the Fourier transform of $f(\br)$ also applies to tight-binding models, which effectively sample potentials on a grid with atomic resolution.}, \cref{eq:Jtilde} can be evaluated as follows on a uniform vertical mesh with step $\delta z=z_{k+1}-z_k=a/4$:
\begin{equation}
J=\langle\tilde{\Phi}^-|H_\mathrm{inter}|\tilde{\Phi} ^+\rangle=\sum_{ijk} \langle\tilde{\Phi}^-_{ijk}|H_\mathrm{inter}(\br_{ijk})|\tilde{\Phi}^+_{ijk}\rangle
\label{eq:Jfd}
\end{equation}
where $\tilde{\Phi}^-_{ijk}=\sqrt{\omega_{ijk}}[\tilde{\varphi}_-(\br_{ijk}), 0]$ and $\tilde{\Phi}^+_{ijk}=\sqrt{\omega_{ijk}}[0, \tilde{\varphi}_+(\br_{ijk})]$ are the discretized wave functions (the total probability amplitudes in the mesh element $\hat{\omega}_{ijk}$ with volume $\omega_{ijk}=(x_{i+1}-x_{i-1})(y_{j+1}-y_{j-1})(z_{k+1}-z_{k-1})/8$), and:
\begin{equation}
H_\mathrm{inter}(\br_{ijk})=s_k V_\mathrm{inter}(\br_{ijk})
\begin{bmatrix}
0 & -i\\
i & 0
\end{bmatrix}
\label{eq:Vinter}
\end{equation}
with $s_k=(-1)^k$ and $V_\mathrm{inter}(\br)\equiv -AV(\br)$. Note that this matrix is imaginary; this follows from the phase conventions for $\tilde{u}_\pm(\br)$ used to write down \cref{eq:h2kpkin}, and has been validated against tight-binding Hamiltonians (see \cref{sec:assumptions}). Practically, we add $H_\mathrm{inter}$ along with $H_\mathrm{VOM}$ to $H_\mathrm{2kp}$, and thus deal with the valley coupling (and possibly valley-orbit mixing) in a variational rather than perturbative way.

The mesh step along $z$ is the distance between successive mono-layers. Therefore, the net inter-valley potential is, up to the prefactor $A$, the intra-valley potential $V$ on ``even'' monolayers, and the opposite of the intra-valley potential on ''odd'' monolayers. In fact, $H_\mathrm{inter}$ is defined up to a global sign, corresponding to the two possible alignments of the atomic lattice with respect to the mesh (namely, $s_k=1$ for monolayers that form bonds to atoms along $[111]$, and $s_k=-1$ for monolayers that form bonds to atoms along $[\bar{1}\bar{1}\bar{1}]$). $H_\mathrm{inter}$ thus changes sign if the whole potential or lattice is shifted by one monolayer -- a well known feature of valley physics.

The mesh along $z$ must, therefore, be very fine (to the monolayer scale), which questions the relevance of this Hamiltonian with respect to fully atomistic tight-binding calculations \cite{Boykin2004,Nestoklon2006,Losert2023}. We emphasize, however, that the meshes along $x$ and $y$ can remain coarse (depending on the scale of variation of the wave function), and that the mesh along $z$ needs to be that fine only in the domain actually probed by the wave function (where valley splitting takes place). This will be illustrated in \cref{sec:qubits}.

\subsection{Strains, spin-orbit coupling, and magnetic fields}

The two bands \kp model can also account for the strains $\varepsilon_{\alpha\beta}$ in the material(s). The strain Hamiltonian is
\begin{equation}
H_\mathrm{2kp}^\mathrm{strains}=
\begin{bmatrix}
H_\mathrm{uniax} & H_\mathrm{shear}  \\
H_\mathrm{shear} & H_\mathrm{uniax} 
\end{bmatrix}
\label{eq:hstrains}
\end{equation}
with:
\begin{subequations}
\begin{align}
H_\mathrm{uniax}&=\Xi_d(\varepsilon_{xx}+\varepsilon_{yy}+\varepsilon_{zz})+\Xi_u\varepsilon_{zz} \\
H_\mathrm{shear}&=2\Xi_s\varepsilon_{xy}\,.
\end{align}
\end{subequations}
Here $\Xi_d$ is the hydrostatic deformation potential, $\Xi_u$ the uniaxial deformation potential, and $\Xi_s$ the shear deformation potential of the conduction band. While hydrostatic and uniaxial strain along $\{100\}$ axes only shift the conduction band edge, shear strains $\varepsilon_{xy}$ mix the $\pm Z$ valleys \cite{Sverdlov2007,Niquet2012,Woods2024,Adelsberger2024}.

Moreover, spin can be introduced in the model by doubling the basis set $\{\tilde{u}_-,\tilde{u}_+\}\to\{\tilde{u}_{-\uparrow},\tilde{u}_{-\downarrow},\tilde{u}_{+\uparrow},\tilde{u}_{+\downarrow}\}$ \cite{Osintsev2014,Osintsev2015,Sverdlov2015}. The resulting four bands \kp Hamiltonian $H_\mathrm{4kp}$ can be straightforwardly derived from \cref{eq:h2kp,eq:h2kpkin,eq:Vinter,eq:hstrains} by replacing each element/operator $m_{ij}$ of the $2\times 2$ matrices by $m_{ij}\otimes I_2$. The bulk Dresselhaus spin-orbit Hamiltonian that splits the spin bands at finite $k_x$ and $k_y$ \cite{Li2011} can then be added to $H_\mathrm{4kp}$. With the spin quantized along $z$,
\begin{equation}
H_\mathrm{4kp}^\mathrm{so}(\bk')=\Delta_\mathrm{so}
\begin{bmatrix}
0_2 & -iD(\bk') \\
iD^\dagger(\bk') & 0_2 
\end{bmatrix}
\label{eq:hso}
\end{equation}
where $D(\bk')=k'_y\sigma_y-k'_x\sigma_x$ (with $\sigma_x$, $\sigma_y$ the Pauli matrices), $0_2$ is the $2\times2$ zero matrix, and $\Delta_\mathrm{so}$ is the Dresselhaus spin-orbit coupling constant. Additional Rashba and Dresselhaus spin-orbit interactions due to symmetry breaking by the interfaces and electric fields may be added the same way if relevant \cite{Golub2004,Ruskov2018,Tanttu2019}.

Finally, the effects of a magnetic field $\vec{B}$ (deriving from a vector potential $\vec{A}$) on the orbital motion can be described by the substitution $-i\vec{\nabla}\to-i\vec{\nabla}+e\vec{A}/\hbar$ in \cref{eq:h2kp} (with $-e$ the electron charge), and its effects on spin by the Zeeman Hamiltonian
\begin{equation}
H_\mathrm{4kp}^\mathrm{zeeman}=\frac{1}{2}g_0\mu_B
\begin{bmatrix}
\vec{B}\cdot\vec{\sigma} & 0_2 \\
0_2 & \vec{B}\cdot\vec{\sigma}
\end{bmatrix}\,,
\label{eq:hzeeman}
\end{equation}
where $g_0\approx2$ is the bare gyromagnetic factor of the electron, $\mu_B$ the Bohr magneton, and $\vec{\sigma}=[\sigma_x,\sigma_y,\sigma_z]$ the vector Pauli matrices.

The parameters of the two and four bands \kp models are the conduction band edge energy $E_c$, the valley wave number $k'_0$, the longitudinal and transverse effective masses $m_l^*$ and $m_t^*$, the deformation potentials $\Xi_d$, $\Xi_u$, $\Xi_s$, and the Dresselhaus parameter $\Delta_\mathrm{so}$. The masses and deformation potentials $\Xi_d$, $\Xi_u$ are well characterized in Si \cite{Niquet2009}, yet not in Ge (as the $Z$ valleys are never the ground-state); and $\Delta_\mathrm{so}$ is not accurately known. We have, therefore recomputed some parameters with the Vienna Ab initio Simulation Package (VASP) \cite{Kresse_1999} using the projector augmented-wave (PAW) method \cite{PAW_1994} and the hybrid HSE functional \cite{HSE_2006} that includes a fraction of exact exchange and provides reasonable band gaps. The parameters used for each material are thus given in Table \ref{tab:parameters}; they are compared with the values calculated with the TB model of Ref.~\cite{Niquet2009}, which will be used to benchmark the two bands \kp model. The parameters for arbitrary Si$_{1-x}$Ge$_x$ alloys are linearly interpolated between those of Si and Ge. 

According to table \ref{tab:parameters}, the strained conduction band offset between Si and a Si$_{0.7}$Ge$_{0.3}$ buffer is $\Delta E_c=166$\,meV in the two bands \kp model, in close agreement with the literature \cite{Schaffler97,Maiti1998}. It is $\Delta E_c=150$\,meV in the TB model \cite{Niquet2009}. The $Z$ valleys of pure Ge are significantly higher in energy in the TB than in the \kp model, but the dependence of the band offset on Ge concentration is also more non-linear, especially at high Ge concentrations. Note that the TB model severely overestimates the bulk Dresselhaus spin-orbit coupling constant $\Delta_\mathrm{so}$ (our {\it ab initio} value for Si is comparable to Ref.~\cite{Li2011} and to all-electron calculations \footnote{The $\Delta_\mathrm{so}$'s were actually computed with the generalized gradients approximation (GGA). We have compared VASP with the all-electron {\it ab initio} code FLEUR \cite{fleurCode} in the local density approximation (LDA) and GGA. All values are in the $1.70-1.85$\,meV.nm range.}).

\begin{table}[!t]
    \centering
    \begin{tabular}{|l|r|r|r|r|}
        \toprule
        & \multicolumn{2}{c|}{Si} & \multicolumn{2}{c|}{Ge} \\
        \cline{2-5}
        & \kp models & TB & \kp models & TB \\
        \hline
        $a$ (\AA) & 5.4298 & 5.431 & 5.6524 & 5.658 \\
        $E_c$ (eV) & 1.170 & 1.173 & 1.500 & 1.644 \\
        $k'_0$ ($2\pi/a$) & 0.159$^\dagger$ & 0.154 & 0.188$^\dagger$ & 0.175 \\
        $m_l^*$ ($m_0$) & 0.916 & 0.900 & 0.89$^\dagger$ & 0.837 \\
        $m_t^*$ ($m_0$) & 0.191 & 0.197 & 0.18$^\dagger$ & 0.178\\
        $\Xi_d$ (eV) & 1.0 & 0.91 & 0.5$^\dagger$ & 0.32 \\
        $\Xi_u$ (eV) & 8.6 & 8.70 & 9.4$^\dagger$ & 9.02 \\
        $\Xi_s$ (eV) & 7.4$^\dagger$ & 8.43 & 8.2$^\dagger$ & 8.11 \\
        $\Delta_\mathrm{so}$ (meV.nm) & 1.7$^\dagger$ & 2.57 & 12.9$^\dagger$ & 19.89 \\
        \botrule
    \end{tabular}
    \caption{The parameters used in the two and four bands \kp models. They are borrowed from the experimental data cited in Refs.~\cite{Niquet2009} and ~\cite{Abadillo2023} or computed {\it ab initio} ($^\dagger$). They are compared to the values calculated with the tight-binding (TB) model of Ref.~\cite{Niquet2009} used to benchmark the \kp models in this work.}
    \label{tab:parameters}
\end{table}

\subsection{About the assumptions of the $2k_0$ theory and the modeling of SiGe alloys}
\label{sec:assumptions}

We can benchmark two bands \kp against atomistic tight-binding (TB) calculations with the first nearest neighbor $sp^3d^5s^*$ model of Ref. \cite{Niquet2009}. For that purpose, we first consider a periodic supercell of bulk silicon with side $a$ along $x=[100]$ and $y=[010]$, and length $L=45a$ along $z=[001]$. We introduce the following potential $V(z)$ that mimics vertical confinement in a Si/SiGe wiggle well:
\begin{subequations}
\label{eq:modelpot}
\begin{align}
V(z)&=0.30E_\mathrm{CBO}\text{ if }|z|>15a\,, \\
V(z)&=0.05E_\mathrm{CBO}\times\frac{1}{2}[1-\cos(qz)]\text{ if }|z|\le 15a\,.
\end{align}
\end{subequations}
The first line indeed describes a quantum well with width $L_\mathrm{QW}=30a\approx 16.3$\,nm and barrier height $\Delta V=0.3E_\mathrm{CBO}$, and the second line a harmonic modulation in the well with wave number $q$ and amplitude $\delta V=0.05E_\mathrm{CBO}$. Here we use $E_\mathrm{CBO}=471$\,meV, the unstrained TB conduction band offset (CBO) between Si and Ge; we also add a periodic electric field $E_z=-5$\,meV/nm in the well and $E_z=10$\,meV/nm in the barrier in order to confine the electrons near the top interface of the well. We deal with this potential in the two bands \kp model as described in \cref{sec:2kp} and \cref{sec:implementation2kp}, and as a macroscopic electric field in the TB Hamiltonian (namely, we add it to the diagonal orbital energies of each atom). We emphasize that we use $V(z)$ as a simple test potential, and do not (yet) intend to describe that way a realistic atomistic or virtual SiGe alloy within TB. We input the TB effective masses of \cref{tab:parameters} in the the two bands \kp model for consistent comparisons.

\begin{figure}
    \centering
    \includegraphics[width=\columnwidth]{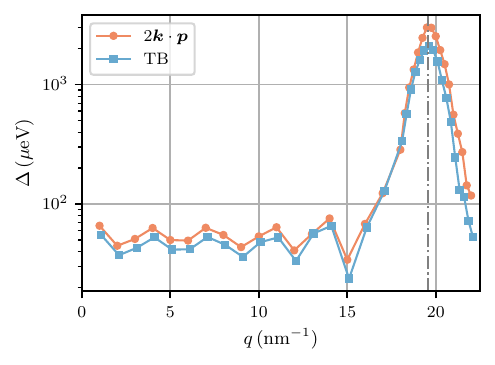}
    \caption{TB and 2 bands \kp valley splittings (2\kp) computed for the model potential of \cref{eq:modelpot} using $A=-0.26$. The vertical dash-dotted line is the inter-valley wave number $q=2k_0$.}
    \label{fig:test}
\end{figure}

The TB and \kp valley splittings computed in this potential are plotted as a function of $q$ in \cref{fig:test}. They are in very good agreement, and show the expected peak at $q=2k_0=19.6$\,nm$^{-1}$ when the modulation in the well is resonant with the inter-valley wave number. The agreement is, in fact, almost perfect if we use $A=-0.225$ instead of $A=-0.26$ in the two bands \kp model. This renormalization is partly due to the fact that the TB Bloch functions are slightly different from the {\it ab initio} Bloch functions used to calculate $A$ with \cref{eq:Acoeff} (see Appendix \ref{app:TB}).

The two bands \kp model (and the underlying $2k_0$ theory) does, therefore, match atomistic TB models very well for simple scalar potentials. We have also compared (as done in \cref{sec:validation}) TB calculations on real, disordered SiGe alloys with two bands \kp calculations that use the local conduction band offset $V_\mathrm{CBO}(\br)$ as potential $V(\br)$ [namely, $V_\mathrm{inter}(\br)=-AV_\mathrm{CBO}(\br)$]. It turns out that the \kp valley splittings are then about four times smaller than the TB valley splittings. We must, therefore, set $|A|\approx 1$ to reproduce the TB data [i.e., $V_\mathrm{inter}(\br)\equiv V_\mathrm{CBO}(\br)$]. This is indeed the prescription made in Ref.~\cite{Losert2023}, but calls for some insights about the assumptions and validity of the $2k_0$ theory.

In fact, the substitution of a Si by a Ge atom can not be accounted for by a slowly varying (thus diagonal) potential in a TB model. As a short-range atomic correction, it modifies, in particular, nearest neighbor (or even longer range) interactions. This is especially relevant for the inter-valley physics, which probes short wavelengths. It is thus necessary to include larger reciprocal lattice vectors $\mathbf{G}-\mathbf{G}'$ in \cref{eq:Jfull} \cite{Thayil2025}. Nevertheless, the structure of \cref{eq:Jinter} suggests that the contributions of Ge atoms in a given ML at $z=z_k$ are essentially additive and proportional to $e^{-2ik_0z_k}|\varphi(z_k)|^2$ (see Appendix \ref{app:Geatoms} for details). \cref{eq:Vinter} may therefore be used with an effective potential $V_\mathrm{inter}$ that accounts for the introduction of Ge atoms and is proportional to the Ge concentration. This is in fact the rationale for the models describing a Ge atom at position $\br_n$ by a delta potential $\lambda\delta(\br-\br_n)$ \cite{Lima2023,Lima2024}. We thus assume
\begin{equation}
V_\mathrm{inter}(\br)=Y(\br)V_\mathrm{inter}^\mathrm{Ge}\,,
\label{eq:Vzk}
\end{equation}
where $Y(\br)$ is the concentration of germanium and $V_\mathrm{inter}^\mathrm{Ge}$ is a parameter. We have adjusted this parameter on TB calculations of the valley splitting in silicon supercells with a single Ge atom (see Appendix \ref{app:Geatoms}), which yield $V_\mathrm{inter}^\mathrm{Ge}\approx 514$\,meV whatever the biaxial strains in the heterostructure. We can compare this parametrization with Ref.~\cite{Losert2023}, which uses the intra-valley conduction band offset $V_\mathrm{inter}(\br)=V_\mathrm{CBO}(\br)$ as inter-valley potential. For SiGe heterostructures grown on a $[001]$-oriented Si$_{1-X}$Ge$_X$ buffer, $V_\mathrm{inter}$ is then, to first-order in $Y$,
\begin{align}
V_\mathrm{inter}&(\br)=Y(\br)[E_c(\mathrm{Si}/\mathrm{Si}_{1-X}\mathrm{Ge}_X) \nonumber \\
&-E_c(\mathrm{Ge}/\mathrm{Si}_{1-X}\mathrm{Ge}_X)]\,,
\end{align}
where $E_c(\mathrm{Si}/\mathrm{Si}_{1-X}\mathrm{Ge}_X)$ is the conduction band edge energy of silicon biaxially strained on Si$_{1-X}$Ge$_X$, and $E_c(\mathrm{Ge}/\mathrm{Si}_{1-X}\mathrm{Ge}_X)$ is the conduction band edge energy of germanium biaxially strained on Si$_{1-X}$Ge$_X$. With the experimental band edge and deformation potentials of \cref{tab:parameters}, $V_\mathrm{inter}^\mathrm{Ge}=E_c(\mathrm{Si}/\mathrm{Si}_{1-X}\mathrm{Ge}_X)-E_c(\mathrm{Ge}/\mathrm{Si}_{1-X}\mathrm{Ge}_X)\approx 570$\,meV over the whole range $X=0-40$\%. Therefore this prescription is, practically, hardly distinguishable from the above parametrization. It is, however, unclear whether $V_\mathrm{inter}(\br)$ is also close to the conduction band offset $V_\mathrm{CBO}(\br)$ in other alloys such as SiC or SiGeC, or if this is a mere coincidence for SiGe heterostructures (as the intra-valley band offset and inter-valley matrix elements have different and {\it a priori} unrelated expressions \footnote{As discussed in Appendix \ref{app:TB}, minimal TB models with only one orbital per atom \cite{Boykin2004} indeed yield $|A|=1$, thus $V_\mathrm{inter}(\br)=V_\mathrm{CBO}(\br)$. This is, however, supported neither by multi-orbital TB models nor by {\it ab initio} calculations \cite{Saraiva2011}.}). We show in the next sections that $V_\mathrm{inter}^\mathrm{Ge}=514$\,meV reliably reproduces the TB valley splittings and inter-valley dipole matrix elements in a wide range of heterostructures of interest for spin qubits. 

In principle, the external (electric field) potential $V_\mathrm{ext}(\br)$ shall still be included in the inter-valley potential using $A\approx-0.26$:
\begin{equation}
V_\mathrm{inter}(\br)=Y(\br)V_\mathrm{inter}^\mathrm{Ge}-AV_\mathrm{ext}(\br)\,.
\end{equation}
The external potential usually makes, nevertheless, little ``direct'' contributions to the valley splitting (it reshapes the envelope functions but has usually no significant Fourier components at $q=2k_0$). Therefore, we discard the external electric field in the inter-valley potential in the following.

\subsection{Alloy disorder}
\label{sec:alloy}

As discussed in Refs.~\cite{PaqueletWuetz2022,Losert2023,Lima2023}, alloy disorder is an essential ingredient of the understanding of valley splittings. In quantum dot systems, alloy disorder scatters the valley splittings, but can enhance their average (since $J$ is a complex number, the ensemble-averaged valley splitting $E[|J|]$ can be larger than $|E[J]|$ if the variance $\sigma_J^2=E[|J-E[J]|^2]$ is large enough).

To deal with alloy disorder on the finite differences mesh, we define an elementary box $\hat{\omega}_{ijk}$ around each mesh point $\br_{ijk}$, with sides $\delta x_i=(x_{i+1}-x_{i-1})/2$, $\delta y_j=(y_{j+1}-y_{j-1})/2$, $\delta z_k=(z_{k+1}-z_{k-1})/2$ and volume $\omega_{ijk}=\delta x_i\delta y_j\delta z_k$ (see \cref{fig:mesh}). We count the total number $N_k=8Y(z_k)\Omega_k/a^3$ of Ge atoms expected in a given monolayer $k$ at $z=z_k$ (with $\Omega_k=\sum_{ij}\omega_{ijk}$) \footnote{$N_k$ actually follows a binomial distribution $B(n=8\Omega_k/a^3, p=Y(z_k))$. We have made the approximation $N_k\approx 8Y(z_k)\Omega_k/a^3$ valid when the area of the simulation box is much larger than the lateral extent of the wave functions.}, and randomly distribute these $N_k$ atoms in all boxes $\hat{\omega}_{ijk}$ of that monolayer. We constrain this distribution so that the number $n_{ijk}$ of Ge atoms in each box does not exceed $n_{ijk}^\mathrm{max}+1$, with $n_{ijk}^\mathrm{max}=8\omega_{ijk}/a^3$ the total number of atoms in that box. We finally define the concentration of Ge atoms $Y_{ijk}=n_{ijk}/n_{ijk}^\mathrm{max}$ in each box and use it in \cref{eq:Vzk} to compute the inter-valley potential $V_\mathrm{inter}(\br_{ijk})$. The boxes $\hat{\omega}_{ijk}$ are practically large enough so that $n_{ijk}^\mathrm{max}>1$ (namely, $\delta x_i\delta y_j>a^2/2$).

In a macroscopically homogeneous alloy with Ge concentration $Y\lesssim 0.5$, the $n_{ijk}$'s follow a binomial distribution $B(n=N_k, p=\omega_{ijk}/\Omega_k)$ with average $np$ and standard deviation $\sigma\approx\sqrt{np}=\sqrt{8Y(z_k)\omega_{ijk}/a^3}$. Therefore, the standard deviation $\sigma/n_{ijk}^\mathrm{max}$ of $Y_{ijk}$ is $\propto \sqrt{a^3/\omega_{ijk}}$: $V_\mathrm{inter}(\br_{ijk})$ looks much noisier on a fine than on a coarse $xy$ mesh. However, the distribution of valley splittings expected from \cref{eq:Jtilde} is little dependent on the mesh as long as the mesh step remains significantly smaller than the extent of the wave function (namely, as long as the wave function probes a statically relevant number of boxes $\hat{\omega}_{ijk}$). In fact, we may expand \cref{eq:Jtilde} on the mesh and compute \cite{Losert2023}
\begin{equation}
E[|J|^2]=|E[J]|^2+\sum_{ijk}\omega_{ijk}^2E[\delta V_\mathrm{inter}^2(\br_{ijk})]|\varphi(\br_{ijk})|^4\,,
\end{equation}
where $\delta V_\mathrm{inter}(\br_{ijk})=V_\mathrm{inter}(\br_{ijk})-E[V_\mathrm{inter}(\br_{ijk})]$. Thus the variances $E[\delta V_\mathrm{inter}^2(\br_{ijk})]$ must indeed be $\propto a^3/\omega_{ijk}$ for this sum to be roughly independent on the mesh steps $\delta x,\,\delta y$ (since $\sum_{ijk}\omega_{ijk}$ is the total volume of the mesh whatever these steps). If the Ge concentration profile only depends on $z$, $\sigma_J^2=E[|J|^2]-|E[J]|^2$ then scales as $(a/r_\parallel)^2$, with $r_\parallel$ the in-plane extension of the wave function \cite{Losert2023}.

Nevertheless, the use of such a rapidly varying potential as an intra-valley potential may break the assumptions of the effective mass theory. We get much better results (with respect to TB) in SiGe heterostructures when we include alloy disorder only in the inter-valley potential $V_\mathrm{inter}$ but not in the intra-valley potential (see Appendix \ref{app:Geatoms} for further justifications). The strains and intra-valley potential are thus computed from the average concentration of germanium $Y(z_k)$ in each monolayer: the lattice parameters, elastic constants, conduction band edge, effective masses and deformation potentials of the Si$_{1-Y}$Ge$_{Y}$ alloy are linearly interpolated between those of Si and Ge (\cref{tab:parameters}) and input in \cref{eq:h2kp} and \cref{eq:hstrains} \footnote{We use symmetric operator ordering $\gamma_i k_\alpha k_\beta\to-\tfrac{1}{2}(\tfrac{\partial}{\partial\alpha}\gamma_i\tfrac{\partial}{\partial\beta}+\tfrac{\partial}{\partial\beta}\gamma_i\tfrac{\partial}{\partial\alpha})$ where the effective masses depend on position.}, while \cref{eq:Vzk} is used for $V_\mathrm{inter}$, with $V_\mathrm{inter}^\mathrm{Ge}=514$\,meV and the above prescription for alloy disorder when relevant.

\section{Validation of the model}
\label{sec:validation}

\begin{figure*}[!t]
  \centering
  \includegraphics{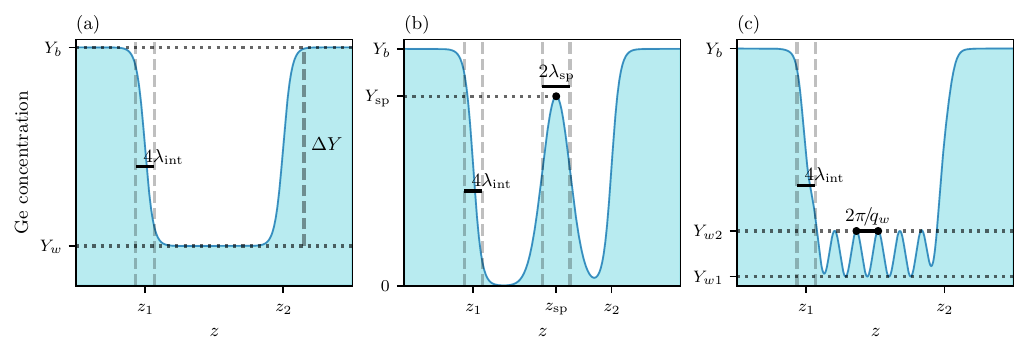}
  \caption{The different Ge concentration profiles considered in this work: (a) Wells with uniform Ge concentration (b) Wells with a Ge spike (c) Wiggle wells.}
  \label{fig:Ge_cprofile}
\end{figure*}

In this section, we validate the above two bands \kp model against atomistic TB calculations on some relevant Si/SiGe heterostructures illustrated in \cref{fig:Ge_cprofile}: SiGe quantum wells with uniform Ge concentration, Si quantum wells with a spike of Ge, and so-called wiggle wells with an oscillatory concentration of Ge \cite{McJunkin2021,Feng2022,McJunkin2022,PaqueletWuetz2022,Losert2023}.

All test structures feature a SiGe quantum well with thickness $W$ and Ge concentration $Y_w$ embedded between a buffer and barrier with Ge concentration $Y_b>Y_w$. The Ge concentration varies smoothly at the interfaces between the well and barriers according to sigmoid functions:
\begin{align}
    Y(z)&=Y_w+\left(Y_b-Y_w\right)\left[\frac{1}{1+e^{(z-z_1)/\lambda_\mathrm{int}}}\right. \nonumber \\ 
    &\left. +\frac{1}{1+e^{-(z-z_2)/\lambda_\mathrm{int}}}\right]\,,
\end{align}
where $z_1=-W/2$ and $z_2=W/2$ are the position of top and bottom interfaces, and $\lambda_\mathrm{int}$ characterizes the diffusion at these interfaces. The actual width of the interfaces is $w_\mathrm{int}\approx 4\lambda_\mathrm{int}$.

The Ge spike \cite{McJunkin2021} is modeled as a Gaussian function:
\begin{equation}
    Y_\mathrm{sp}(z)=Y_\mathrm{sp}e^{-(z-z_\mathrm{sp})^2/(2\lambda_\mathrm{sp}^2)}\,,
    \label{eq:Gespike}
\end{equation}
where $Y_\mathrm{sp}$ is the concentration of the spike, $z_\mathrm{sp}$ its position and $\lambda_\mathrm{sp}$ characterizes its width.

Finally, wiggle wells \cite{Feng2022,McJunkin2022} are modeled by the following oscillatory profile:
\begin{equation}
    Y_w=Y_{w1}+\frac{Y_{w2}-Y_{w1}}{2}\left[1-\cos(qz)\right]\,,
    \label{eq:Geww}
\end{equation}
where $Y_{w1}$ and $Y_{w2}$ are the minimum and maximum Ge concentration in the wiggle well, respectively, and $q$ is the wave number of the oscillations.
 
In the following, the thickness of the well is $W=80$\,ML ($\approx 11$\,nm). We apply periodic boundary conditions over a supercell of length $l_z=160$\,ML and $l_x=l_y=44$\,nm. The electron is confined in the $xy$ plane by a harmonic potential such that the radius of the dot is $r_\parallel\approx 9$\,nm and the orbital splitting is $\Delta_\mathrm{orb}\approx 5$\,meV. We can, moreover, apply a vertical electric field $E_z$ in the well and $-E_z$ in the barrier to squeeze the electron at the top interface \footnote{The electric field is opposite in the well and barrier (sawtooth potential) in order to meet periodic boundary conditions over the supercell.}. The SiGe alloys are modeled as random distributions of Si and Ge atoms (achieving the target Ge concentration $Y(z)$ in each monolayer) in TB, and as described in \cref{sec:alloy} in \kp calculations. The structures are biaxially strained onto the buffer. The atomic positions are relaxed with Keating's valence force field for TB \cite{Keating1966, Niquet2009}. The strains $\varepsilon_{\alpha\beta}$ used as input for the intra-valley potential [\cref{eq:hstrains}] are calculated with continuum elasticity theory (neglecting alloy disorder \footnote{Namely, $\varepsilon_{zz}=-2c_{12}/c_{11}\varepsilon_\parallel$, where $c_{11}$ and $c_{12}$ are the bulk elastic constants, and $\varepsilon_{xx}=\varepsilon_{yy}=\varepsilon_\parallel=(a-a_b)/a$ are the in-plane strains, with $a$ the lattice parameter of the Si$_{1-Y}$Ge$_Y$ material at a given $z$ and $a_b$ the lattice parameter of the buffer. The elastic constants and lattice parameter of Si$_{1-Y}$Ge$_Y$ are linearly interpolated between those of pure Si and Ge.}) and the elastic constants of Ref.~\cite{Abadillo2023}. We use the experimental/{\it ab initio} parameters of \cref{tab:parameters} in the two bands \kp model.

The mesh of the \kp calculations is uniform with steps $\delta x=\delta y=1$\,nm and $\delta z=a/4$ (619\,520 degrees of freedom). The TB supercells contain 2\,048\,000 atoms and 20\,480\,000 orbitals. Each two bands \kp calculation (for a given concentration profile and realization of the disorder) is on average $\approx 250\times$ faster than the corresponding TB calculation.

\subsection{Wells with uniform Ge concentration}

\begin{figure}
    \centering
    \includegraphics[width=\columnwidth]{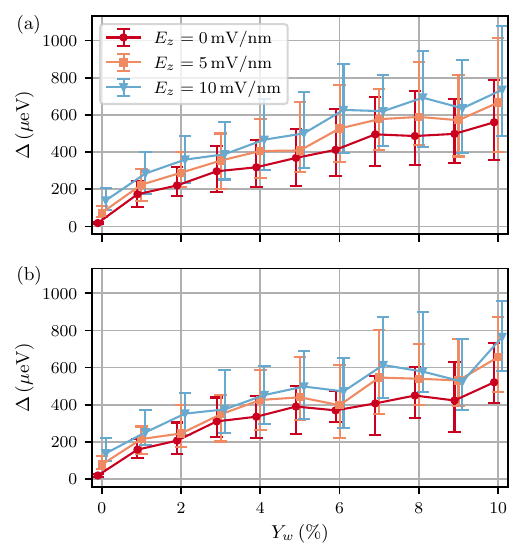}
    \caption{(a) Two bands \kp and (b) TB valley splittings in a quantum well with uniform Ge concentration $Y_w$ and constant $\Delta Y=Y_b-Y_w=30$\%, for different vertical electric fields $E_z$. The median splittings, computed over 64 (TB) or 128 (\kpns) disorder configurations, are plotted with error bars giving the inter-quartile range. The interface width is $w_\mathrm{int}=10$\,ML.}
    \label{fig:vs_unifw10}
\end{figure}

\begin{figure}
    \centering
    \includegraphics[width=\columnwidth]{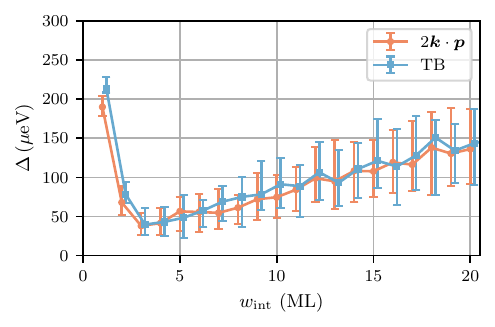}
    \caption{Two bands \kp and TB valley splittings in a plain Si quantum well ($Y_w=0$), as a function of the interface width $w_\mathrm{int}$. The median splittings, computed over 64 (TB) or 128 (\kpns) disorder configurations, are plotted with error bars giving the inter-quartile range. The vertical electric field is $E_z=5$\,mV/nm.}
    \label{fig:vs_smwintdot}
\end{figure}

As a first example, we consider quantum wells with residual Ge concentration $Y_w$ and constant $\Delta Y=Y_b-Y_w=30$\% \cite{PaqueletWuetz2022}. The interface width is $w_\mathrm{int}=10$\,ML. The TB and \kp valley splittings are plotted as a function of $Y_w$ in \cref{fig:vs_unifw10}, for different vertical electric fields $E_z$. The median valley splitting (lines and dots) as well as the inter-quartile range (error bars) are given. These statistics are more robust against outliers than the average and standard deviation. The agreement between the two methods is very good. We emphasize that the trends shown by this figure are dominated by alloy disorder. Indeed, the Ge concentration profile itself does not display a large Fourier component at $q=2k_0$, but the alloy disorder within the well may do so (although the magnitude and phase of the resulting $J$ are random). As discussed in Ref. \cite{Losert2023}, $J$ is expected to follow a Gaussian distribution in the complex plane, and $|J|$ a Rician distribution whose average $E[|J|]$ is greater than $|E[J]|$ when the variance $\sigma_J^2=E[|J-E[J]|^2]$ is large. There is, nevertheless, a tail of devices with valley splittings down to zero in the distribution. As expected, the median valley splitting increases with the residual concentration $Y_w$ (which strengthens the disorder), and with the electric field (as the wave function further probes the smooth top interface with larger Ge concentration). We also plot the valley splitting in a plain Si well ($Y_w=0$) as a function of the interface width $w_\mathrm{int}$ in \cref{fig:vs_smwintdot} ($E_z=5$\,mV/nm). It shows a minimum around $w_\mathrm{int}\approx3$\,ML \cite{Lima2023,Losert2023}. Indeed, the valley splitting is dominated by the $2k_0$ component of the sharp barrier when $w_\mathrm{int}\to 0$, and by the alloy disorder in the smoothed interface at large $w_\mathrm{int}$. Such an enhancement of the valley splitting by broad interfaces has been experimentally demonstrated in Ref. \cite{Stehouwer2025}. The TB and 2 bands \kp model are again in good agreement. The data shown in \cref{fig:vs_unifw10} and \cref{fig:vs_smwintdot} are qualitatively consistent with those of Ref. \cite{Losert2023} (however calculated with different structural parameters).

\subsection{Ge spike in the well} 
\label{sec:spikes}

\begin{figure}
    \centering
    \includegraphics[width=\columnwidth]{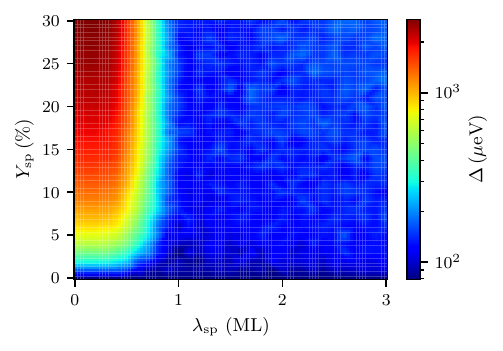}
    \caption{Two bands \kp valley splitting in a Si quantum well as a function of the width $\lambda_\mathrm{sp}$ and concentration $Y_\mathrm{sp}$ of a Ge spike (median of 64 disorder configurations). The barrier concentration is $Y_b=30$\%, the interface width $w_\mathrm{int}=10$\,ML, and the spike is centered at $z_\mathrm{sp}=5$\,ML above the middle of the well. The vertical electric field is $E_z=5$\,mV/nm.}
    \label{fig:vs_spike3D}
\end{figure}

\begin{figure}
    \centering
    \includegraphics[width=\columnwidth]{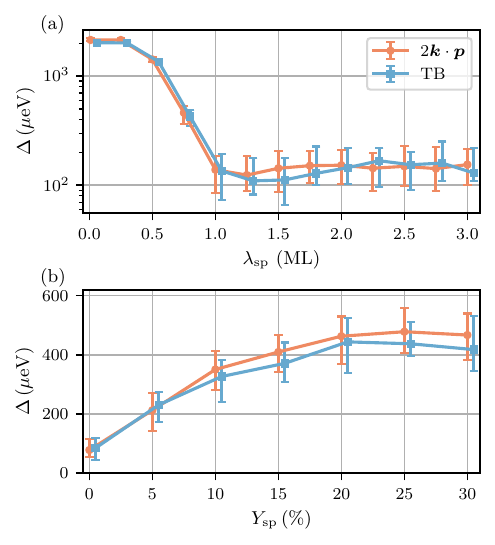}
    \caption{Two bands \kp and TB valley splittings in a Si quantum well with a Ge spike (a) as a function of $\lambda_\mathrm{sp}$ for $Y_\mathrm{sp}=20$\% and (b) as a function of $Y_\mathrm{sp}$ for $\lambda_\mathrm{sp}=0.75$\,ML. The median splittings, computed over 64 (TB) or 128 (\kpns) disorder configurations, are plotted with error bars giving the inter-quartile range. The other structural parameters are the same as in  \cref{fig:vs_spike3D}, and the vertical electric field is $E_z=5$\,mV/nm.}
    \label{fig:vs_spike2Dcomb}
\end{figure}

We next consider a spike of Ge in the well \cite{McJunkin2021}, with a Gaussian profile defined by \cref{eq:Gespike}. We set $Y_w=0$, $Y_b=30$\%, $w_\mathrm{int}=10$\,ML, and center the spike $z_\mathrm{sp}=5$\,ML above the middle of the well. The median two bands \kp valley splitting is plotted as a function of the spike concentration $Y_\mathrm{sp}$ and spike width $\lambda_\mathrm{sp}$ in \cref{fig:vs_spike3D} ($E_z=5$\,mV/nm). The Fourier transform of this Gaussian spike is a Gaussian with amplitude $\propto Y_\mathrm{sp}$ and width $\propto 1/\lambda_\mathrm{sp}$; the median valley splitting thus increases with increasing $Y_\mathrm{sp}$ and decreasing $\lambda_\mathrm{sp}$. As a matter of fact, the largest valley splittings achieved when $\lambda_\mathrm{sp}\lesssim 1$\,ML are ruled by the $q_z=2k_0$ Fourier component of the spike rather than by disorder (but growing such thin spikes remains challenging). Note, however, that the data reported in this figure are actually the splitting between the lowest two levels of the quantum dot; they thus saturate around $\Delta\simeq 5$\,meV when the second level becomes an orbital instead of a valley excitation. These trends are, again, consistent with Ref. \cite{Losert2023}.

We compare the two bands \kp and TB data on selected lines of this map in \cref{fig:vs_spike2Dcomb} (as a function of $\lambda_\mathrm{sp}$ for fixed $Y_\mathrm{sp}=20$\%, and as a function of $Y_\mathrm{sp}$ for fixed $\lambda_\mathrm{sp}=0.75$\,ML). The overall agreement between the two models is very satisfactory.

\subsection{Wiggle wells}

\begin{figure}
    \centering
    \includegraphics[width=\columnwidth]{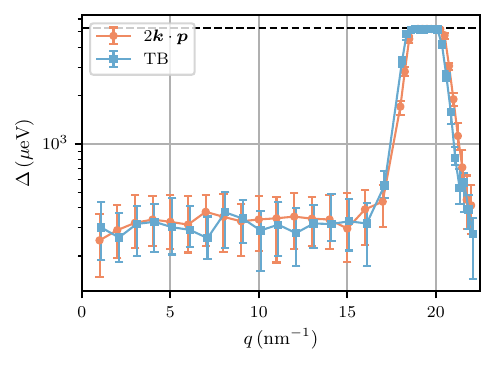}
    \caption{Two bands \kp and TB valley splittings in a wiggle well as a function of the wave number $q$ of the modulation. The median splittings, computed over 64 (TB) or 128 (\kpns) disorder configurations, are plotted with error bars giving the inter-quartile range. The horizontal dashed line is the orbital splitting $\Delta_\mathrm{orb}=5.3$\,meV. The barrier concentration is $Y_b=30$\%, the interface width $w_\mathrm{int}=10$\,ML, and the electric field is $E_z=5$\,mV/nm.}
    \label{fig:wiggle_3DTB}
\end{figure}

Finally, we consider wiggle wells \cite{Feng2022,McJunkin2022} with oscillating Ge concentration [\cref{eq:Geww}]. We set $Y_b=30$\%, $Y_{w1}=0$, $Y_{w2}=5$\%, and $w_\mathrm{int}=10$\,ML. The two bands \kp and TB valley splittings are plotted as a function of the wave number $q$ of the oscillations in \cref{fig:wiggle_3DTB} ($E_z=5$\,mV/nm). As expected, the valley splitting peaks when $q\approx2k_0=19.6$\,nm$^{-1}$. The peaks are, however, cut above $\Delta\approx 5$\,meV as the valley splitting becomes larger than the orbital splitting (see previous subsection). The TB and two bands \kp data are, once again, in good agreement (also on the energy of the orbital excitation). The TB and \kp peaks are slightly shifted one with respect to the other due to slightly different $k_0$'s (see \cref{tab:parameters}).

We conclude from these comparisons that the present two bands \kp model provides a faithful description of the valley splittings in the Si/SiGe heterostructures of interest for spin qubits.

\section{Valley-orbit mixing and dipole matrix elements} 
\label{sec:voc_dip}

\begin{figure}
    \centering
    \includegraphics[width=\columnwidth]{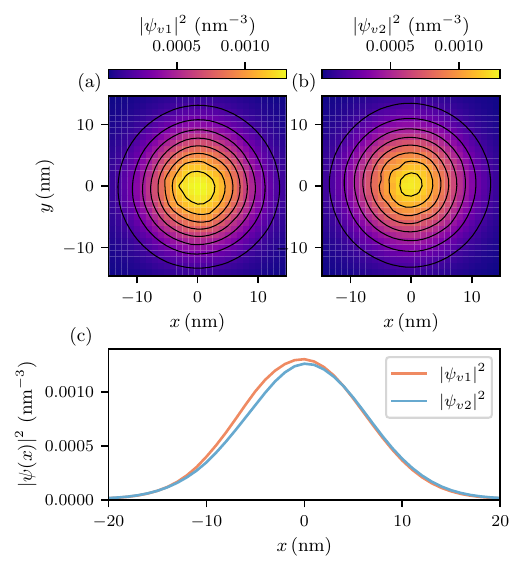}
    \caption{Electron densities $|\psi|^2\equiv|\tilde{\varphi}_+|^2+|\tilde{\varphi}_-|^2$ in a quantum well with a spike of Ge ($Y_\mathrm{sp}=20$\%, $\lambda_\mathrm{sp}=0.75$\,ML with alloy disorder). (a)-(b) Densities of the two ground valley states $|\psi_{v1}|^2$ and $|\psi_{v2}|^2$ in the $xy$ plane at $z=z_\mathrm{max}=\operatorname{arg\,max}_z |\psi_{v1}(x=0,y=0,z)|^2$, for a particular realization of disorder. (c) Densities of the two ground valley states along $x$ for $(y,z)=(0,z_\mathrm{max})$, for the same realization of disorder. The other structural parameters are the same as in \cref{fig:vs_spike3D} and the vertical electric field is $E_z=5$\,mV/nm.}
    \label{fig:wf_sq}
\end{figure}

In this section, we show that the two-bands \kp model also describes valley-orbit mixing effects accurately.

In the structures considered in the previous section, the potential is separable along $x$, $y$ and $z$. In the absence of disorder, the wave functions $\psi_{v1}$ and $\psi_{v2}$ of the two valley states can thus be written as the product of the same in-plane envelope $\phi^\parallel(x,y)$ and of different, orthogonal out-of-plane wave functions $\psi_{v1,v2}^\perp(z)$. The inter-valley dipole matrix elements $\langle\psi_{v2}|x|\psi_{v1}\rangle$ and $\langle\psi_{v2}|y|\psi_{v1}\rangle$ are thus zero. The motions along $x$, $y$ and $z$ get however coupled by alloy disorder, which results in finite dipole matrix elements. This can alternatively be analyzed as an effect of valley-orbit mixing (the admixture of different orbitals by the inter-valley potential) \cite{Gamble2013}. The wave functions of the two valley states become visibly different not only along $z$ but also in any $xy$ plane, as illustrated in \cref{fig:wf_sq}. As discussed in the next section, the existence of a finite inter-valley dipole can leave visible fingerprints in the dynamics of spin and valley qubits.

\begin{figure}
    \centering
    \includegraphics[width=\columnwidth]{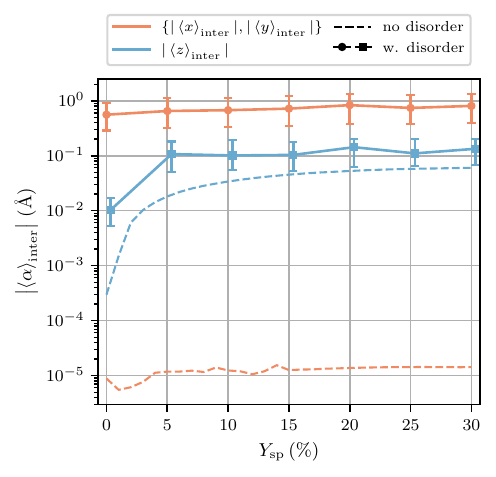}
    \caption{Two bands \kp inter-valley dipole matrix elements in a Si quantum well as a function of the concentration $Y_\mathrm{sp}$ of a Ge spike, with (solid lines) and without (dashed lines) alloy disorder. The width of the spike is $\lambda_\mathrm{sp}=0.75$ ML and the other structural parameters are the same as in \cref{fig:vs_spike3D} (in particular, $w_\mathrm{int}=10$\,ML). The vertical electric field is $E_z=5$\,mV/nm. With disorder, the median values, computed over 128 configurations, are plotted with error bars giving the inter-quartile range. $|\langle x\rangle_\mathrm{inter}|$ and $|\langle y\rangle_\mathrm{inter}|$ have the same statistics and have been lumped together (or averaged in the undisordered device).}
    \label{fig:dip_sp_NDDcomp}
\end{figure}

In order to assess the effects of valley-orbit mixing, we monitor, therefore, the inter-valley dipole matrix elements 
\begin{equation}
\langle\alpha\rangle_\mathrm{inter}=\langle\psi_{v2}|\alpha|\psi_{v1}\rangle
\end{equation}
where $\alpha\in\{x,\,y,\,z\}$. The statistics of the two bands \kp dipole matrix elements in a Si well with a Ge spike are plotted a function of $Y_\mathrm{sp}$ in \cref{fig:dip_sp_NDDcomp}, with and without alloy disorder (same structures as in \cref{sec:spikes} with $\lambda_\mathrm{sp}=0.75$\,ML). As expected, $\langle x\rangle_\mathrm{inter}$ and $\langle y\rangle_\mathrm{inter}$ are (numerically) small in the absence of alloy disorder. There is, however, a finite $\langle z\rangle_\mathrm{inter}$ resulting from the out-of-phase oscillations of the real and imaginary parts of the valley envelopes at wave number $q_z\approx k_0'$. This vertical dipole is visibly enhanced by the presence of the spike, although it saturates rapidly with increasing $Y_\mathrm{sp}$. The picture changes drastically when alloy disorder is taken into account. The in-plane dipoles $|\langle x\rangle_\mathrm{inter}|$ and $|\langle y\rangle_\mathrm{inter}|$ now supersede $|\langle z\rangle_\mathrm{inter}|$. Indeed, the disordered inter-valley potential effectively structures the valley envelopes in the $xy$ plane since the orbital splittings are smaller for in-plane than for vertical excitations. We emphasize that this in-plane dipole is randomly oriented (as the disorder does not break, on average, the symmetry of the dot). Therefore, $E[\langle x\rangle_\mathrm{inter}]$ and $E[\langle y\rangle_\mathrm{inter}]$ remain zero even though $E[|\langle x\rangle_\mathrm{inter}|]=E[|\langle y\rangle_\mathrm{inter}]|$ are finite. The analysis of the perturbation series for $\langle x\rangle_\mathrm{inter}$ and $\langle y\rangle_\mathrm{inter}$ suggests that $\sqrt{E[\langle x\rangle_\mathrm{inter}^2]}$ and $\sqrt{E[\langle y\rangle_\mathrm{inter}^2]}$ scale as $r_\parallel^2/a$, which has been verified numerically.

\begin{figure}
    \centering
    \includegraphics{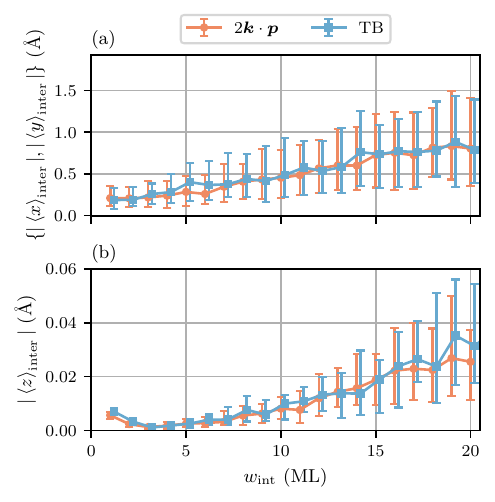}
    \caption{Two bands \kp and TB inter-valley dipole matrix elements in a plain Si quantum well ($Y_w=0$), as a function of the interface width $w_\mathrm{int}$, with alloy disorder. The vertical electric field is $E_z=5$\,mV/nm. The median values, computed over 64 (TB) or 128 (\kpns) disorder configurations, are plotted with error bars giving the inter-quartile range.}
    \label{fig:dip_smwint}
\end{figure}

\begin{figure}
    \centering
    \includegraphics{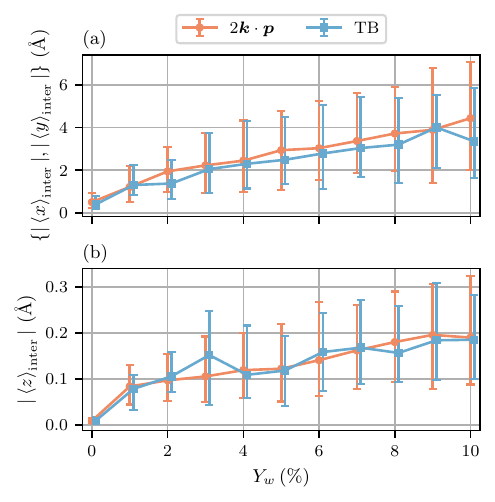}
    \caption{Two bands \kp and TB inter-valley dipole matrix elements in a Si quantum well with uniform Ge concentration $Y_w$ and constant $\Delta Y=Y_b-Y_w=30$\%, with alloy disorder. The interface width is $w_\mathrm{int}=10$\,ML and the vertical electric field is $E_z=5$\,mV/nm. The median values, computed over 64 (TB) or 128 (\kpns) disorder configurations, are plotted with error bars giving the inter-quartile range.}
    \label{fig:dip_smYw}
\end{figure}

The dipoles $|\langle x\rangle_\mathrm{inter}|$ and $|\langle y\rangle_\mathrm{inter}|$ show a weak dependence on $Y_\mathrm{sp}$, which suggests that they primarily result from disorder at the smooth Si/SiGe interfaces (for the present structures). We thus plot the two bands \kp and TB dipole matrix elements in plain Si quantum wells as a function of the interface width $w_\mathrm{int}$ in \cref{fig:dip_smwint} (with the same parameters as in \cref{fig:vs_smwintdot}). The in-plane dipoles at $w_\mathrm{int}=10$\,ML are indeed comparable to \cref{fig:dip_sp_NDDcomp}, but $|\langle z\rangle_\mathrm{inter}|$ is much smaller (and consistent with the limit $Y_\mathrm{sp}\to0$). Both the in-plane and vertical dipoles steadily grow with $w_\mathrm{int}>3$\,ML as the wave functions penetrate farther into the disordered alloy. The agreement between the TB and two bands \kp models is very good, which demonstrates that the latter captures valley-orbit mixing accurately.

Finally, we compare the two bands \kp and TB dipole matrix elements in quantum wells with a residual Ge concentration $Y_w$ in \cref{fig:dip_smYw} ($\Delta Y=Y_b-Y_w=30$\% and interface width $w_\mathrm{int}=10$\,ML as in \cref{fig:vs_unifw10}). As expected, the average inter-valley dipole matrix elements increase with alloy disorder in the well (as does the valley splitting $\Delta$). The in-plane dipole matrix elements can, in particular, be significant down to small $Y_w\gtrsim 2\%$. The agreement between TB and the two bands \kp model is, once again, very satisfactory. 

\section{Application to spin qubits}
\label{sec:qubits}

We next illustrate the application of the two bands \kp model to spin qubits.

\subsection{Device}

\begin{figure}
    \centering
    \includegraphics[width=.9\columnwidth]{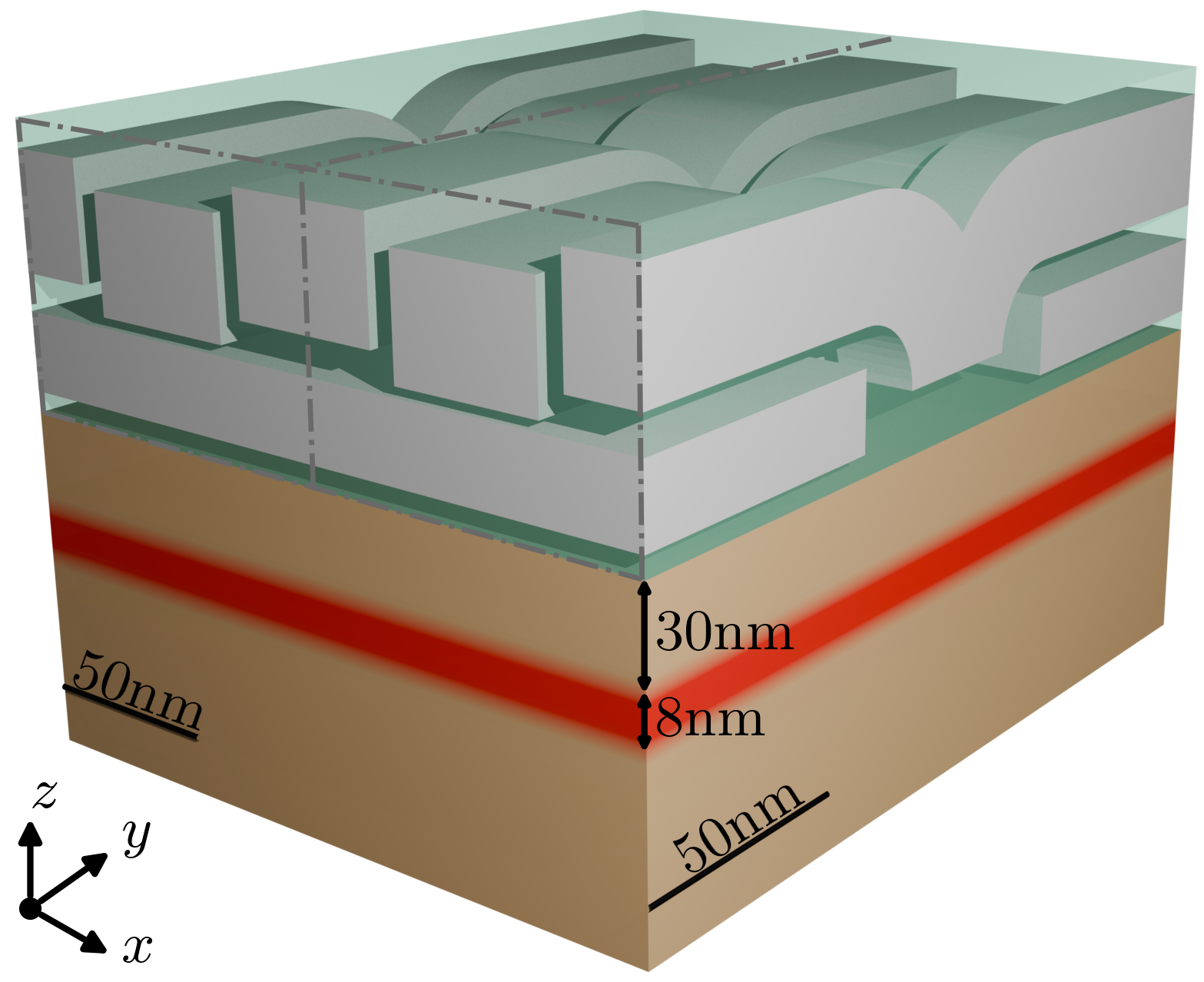}
    \caption{Three-dimensional representation of the qubit device considered for the simulations. The Si$_{0.7}$Ge$_{0.3}$ buffer and barrier are shown in brown, the silicon well in red (with a red/brown gradient at the smoothed interfaces), the gates in gray and the aluminium oxide in green. The dashed gray lines outline the two cross sections shown in \cref{fig:xzyztopqubit}. The origin is at the center of the quantum well.}
    \label{fig:qbit3d}
\end{figure}

\begin{figure}
    \centering
    \includegraphics[width=\columnwidth]{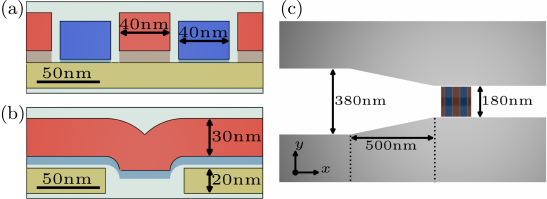}
    \caption{(a) $xz$ cross section at $y=-90$\,nm and (b) $yz$ cross section at $x=0$ of the gate stack of the device of \cref{fig:qbit3d}. Plunger gates are shown in red, barrier gates in blue and screening gates in yellow. (c) Top view of the qubit device and micro-magnet used to drive Rabi oscillations electrically. Each half of the micro-magnet is 2\,$\mu$m long along $x$ and 1.5\,$\mu$m wide along $y$. The micro-magnet is 200\,nm thick. The two cross sections in (a) and (b) are outlined in \cref{fig:qbit3d}.}
    \label{fig:xzyztopqubit}
\end{figure}

The qubit device considered in these simulations is inspired by Ref.~\cite{Philips2022} and is shown in \cref{fig:qbit3d}. We plot cross sections of this device in \cref{fig:xzyztopqubit}, as well as a top view showing the design of the micro-magnet used to enable electric dipole spin resonance (EDSR). 

The heterostructure is made of a $60\,\mathrm{ML}\approx 8.1$\,nm thick Si well lattice matched to a Si$_{0.7}$Ge$_{0.3}$ buffer and capped with a 30\,nm thick Si$_{0.7}$Ge$_{0.3}$ barrier. There are three layers of gates (identified by different colors), separated by 5\,nm thick Al$_2$O$_3$ oxides. In the first level, two 20\,nm thick screening gates (yellow), split by a 60\,nm wide gap, confine the carriers in a one-dimensional channel along $x$. Plunger gates (red, third level) and barrier gates (blue, second level) shape quantum dots in this channel and control their interactions. They are 40\,nm wide and 30\,nm thick. The side of the simulation box along $x$ and $y$ is $L=180$\,nm. The in-plane mesh steps are $\delta x=\delta y=1$\,nm below the plunger gate, and are made gradually larger under the barrier and screening gates. The out-of-plane mesh step is $\delta z=a/4$ for $|z|<12.2$\,nm ($z=0$ being the middle of the well). The wave functions are assumed to be zero (and the $z$ mesh gets coarser) outside this domain.

The synthetic spin-orbit interaction required for electrical spin manipulation is provided by the inhomogeneous field of a Co micro-magnet deposited $100$\,nm above the heterostructure (see \cref{fig:xzyztopqubit}c). The whole device is placed in an external magnetic field $\vec{B}=B_y\hat{\vec{y}}$ that saturates the polarization $J_m=1.84$\,T of this micro-magnet \cite{Neumann2015}. 

The field of the micro-magnet is calculated as in Ref.~\cite{Martinez2022}. We find magnetic field distributions very close to \cite{Philips2022}. The potential $V(\br)$ in the device is computed on the mesh with a finite-volume solver for Poisson's equation
\begin{equation}
\boldsymbol{\nabla}_{\vec{r}}\cdot\kappa(\vec{r})\boldsymbol{\nabla}_{\vec{r}}V(\vec{r})=0\,,
\end{equation}
with $\kappa(\vec{r})=\kappa_0\kappa_r(\vec{r})$, $\kappa_0$ the vacuum permittivity and $\kappa_r(\vec{r})$ the dielectric constant at point $\vec{r}$ \footnote{The dielectric constants are $\kappa_r=11.7$ for Si, $\kappa_r=16.2$ for Ge, and $\kappa_r=8$ for Al$_2$O$_3$.}. The bias voltages are used as boundary conditions on the gates, while zero normal electric field (Neumann boundary conditions) is imposed on the edges of the simulation box \cite{Martinez2022hole}. The spin-valley physics is described by the four bands \kp Hamiltonian, discretized using finite differences on the mesh. The bulk Dresselhaus interaction is included, but is practically negligible with respect to the synthetic spin-orbit field. As discussed above, the wave function amplitudes, which decrease exponentially in the barriers, are negligible when $|z|\gtrsim 10$\,nm. The \kp equations are, therefore, solved in the domain $|z|<12.2$\,nm.

In order to achieve sizable valley splittings, we consider either a quantum well with a residual Ge concentration $Y_w=2$\% or a Ge spike ($Y_\mathrm{sp}=20$\%, $\lambda_\mathrm{sp}=1$\,ML, $z_\mathrm{sp}=5$\,nm). The interface width is $w_\mathrm{int}=4$ ML in both cases, and alloy disorder is accounted for. We ground the screening and barrier gates and bias the central plunger gate at $V_P=0.2$\,V. We reach a valley splitting $\Delta=202\,\mu$eV for the alloyed well and $\Delta=115\,\mu$eV for the Ge spike with the random alloy configuration chosen for this study. This configuration is representative of the valley splittings achieved with each concentration profile ({\it i.e.}, is not an outlier), but we will not address statistics in this section, as they have been extensively discussed in \cref{sec:validation,sec:voc_dip}.

\subsection{Rabi Frequency, relaxation and dephasing times}

We consider the qubit based on the lowest two states $\ket{\zero}$ and $\ket{\one}$ of the dot. At low magnetic field, $\ket{\zero}$ and $\ket{\one}$ are the spin states $\ket{v_1,\downarrow}$ and $\ket{v_1,\uparrow}$ of the ground-state valley, split by the energy $h f_L \approx g^*\mu_B B_y$, where $g^*\approx 2$ is the effective $g$-factor of the electron (see \cref{fig:frTsvsB}a). At high magnetic field, they are the spin down valley states $\ket{v_1,\downarrow}$ and $\ket{v_2,\downarrow}$ split by $h f_L\approx\Delta$. The crossover between the spin and valley qubit regimes takes place at the magnetic field $B_c=\Delta/g^*\mu_B$ (see later discussion). 

We resonantly drive this qubit with an AC signal $V_\mathrm{sc}(t)=\delta V_\mathrm{ac}\sin(2\pi f_L t+\varphi)$ applied to one of the two screening gates, thus shaking the dot along the $y$ axis. In both the spin and valley qubit regimes, the Rabi frequency is, to first-order in $\delta V_\mathrm{ac}$,
\begin{equation}
    f_R=\frac{e}{h}\delta V_\mathrm{ac}\left|\langle\one|D_\mathrm{sc}|\zero\rangle\right|\,,
\end{equation}
where $D_\mathrm{sc}(\br)=\partial V(\br)/\partial V_\mathrm{sc}$ is the electric dipole operator (the linear response of the total electrostatic potential $V(\bm{r})$ to a change of the screening gate potential $V_\mathrm{sc}$). If the electrostatics is linear (which is the case in the absence of dense electron gases around), $D_\mathrm{sc}(\br)$ is simply the potential created by a pulse of 1\,V on the screening gate with all other gates grounded. 

We also compute the relaxation rate $\Gamma_1$ due to electron-phonon interactions. We assume bulk-like acoustic phonons and account for the interplay between intra-valley scattering (characterized by the hydrostatic and uniaxial deformation potentials $\Xi_d$ and $\Xi_u$) and inter-valley scattering (characterized by the shear strain deformation potential $\Xi_s$) \cite{Li2020}. The effects of the phonon-induced shear strains are usually neglected in the literature; we discuss below in which regimes they can be relevant. The equations for the relaxation rate including all deformation potentials are given in Appendix \ref{app:eph}.

Finally, we give some estimates of the electrical dephasing time $T_2^*=(\Gamma_2^*)^{-1}$ in a simple approximation. We assume as in Ref. \cite{Mauro2024} that the different sources of electrical $1/f$ noise can be lumped into effective gate voltage variations. Then, 
\begin{equation}
    \Gamma_2^*=\sqrt{2}\pi\sqrt{\sum_i\left(\delta V_i\frac{\partial f_L}{\partial V_i}\right)^2}\,,
    \label{eq:dephrate}
\end{equation}
where $\partial f_L/\partial V_i=e\left(\langle\one|D_i|\one\rangle-\langle\zero|D_i|\zero\rangle\right)/h$ and the index $i$ runs over the gates. We consider the same noise amplitude $\delta V_i=\delta V=10\,\mu$eV on all gates.

\subsection{Results and discussion}

\begin{figure}
    \centering
    \includegraphics[width=\columnwidth]{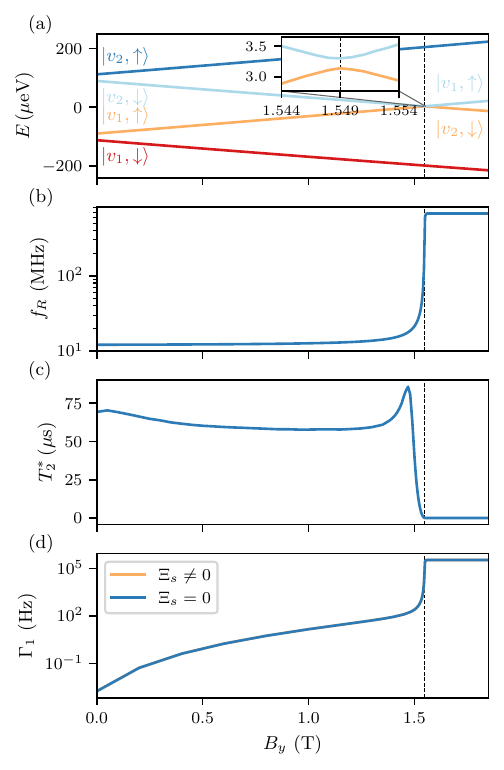}
    \caption{Spin qubit in a quantum well with 2\% residual Ge. (a) Energy levels of the qubit as a function of the external magnetic field $B_y$ along $y$. The Zeeman splitting does not vanish when $B_y\to0$ due to the  magnetic field of the micro-magnet. (b) Rabi frequency $f_R$ as a function of $B_y$ (driving a screening gate with a signal of amplitude $\delta V_\mathrm{ac}=1$\,mV). (c) Dephasing time $T_2^*$ as a function of $B_y$ assuming noise amplitudes $\delta V_i=10\,\mu$V on all gates. (d) Spin-phonon relaxation rate $\Gamma_1$ calculated with and without the shear strain deformation potential $\Xi_s$, at temperature $T=100$\,mK. The $\Xi_s\ne0$ curve (orange line) is not distinguishable from the $\Xi_s=0$ curve (blue line) in the present case. The vertical dashed line in all panels indicates the magnetic field at which the Zeeman energy exceeds the valley splitting. The device becomes a valley qubit past this critical field.}
    \label{fig:frTsvsB}
\end{figure}

The Rabi frequency $f_R$, dephasing time $T_2^*$ and relaxation rate $\Gamma_1$ in the well with 2\% residual Ge are plotted as a function of the external magnetic field $B_y$ in \cref{fig:frTsvsB}. The transition from the spin qubit to the valley qubit regime at $B_y=1.549$\,T is well visible on all plots ($\Delta=202\,\mu$eV). The Rabi frequency is $f_R\approx 12$\,MHz in the spin qubit regime for drive amplitude $\delta V_\mathrm{ac}=1$\,mV. This is consistent with the displacement of the driven dot in the gradient of magnetic field. Indeed, when spin-orbit interactions are dominated by a micro-magnet, the Rabi frequency shall be (far from the spin/valley anti-crossing) \cite{Martinez2022}
\begin{equation}
f_R=\frac{g_0\mu_B}{2h}\left|\vec{b}\times\left(G \delta\langle\br\rangle\right)\right|\,,
\end{equation}
where $\vec{b}$ is the unit vector along the average magnetic field in the dot, $G$ is the matrix of the average magnetic field gradients in the dot ($G_{ij}=\partial B_i/\partial r_j$), and $\delta\langle\br\rangle$ is the displacement of the driven dot at $V_\mathrm{sc}=\delta V_\mathrm{ac}$. In the present device, the $G$ matrix reads at the origin (the center of the well):
\begin{equation}
G=\begin{pmatrix}
\approx 0 & -0.041 & \approx 0 \\
-0.041 & \approx 0 & 1.88 \\
\approx 0 & 1.88 & \approx 0 \\
\end{pmatrix}\,\text{mT/nm}\,.
\end{equation}
Note that this matrix is symmetric and traceless as $\boldsymbol{\nabla}\cdot\vec{B}=0$ and $\boldsymbol{\nabla}\times\vec{B}=\vec{0}$ outside the magnet. Moreover, $\vec{b}\approx\hat{\vec{y}}$ and $\delta\langle\br\rangle/\delta V_\mathrm{ac}\approx0.46$\,nm/mV along $\hat{\vec{y}}$ so that the expected Rabi frequency is $f_R\approx (g_0\mu_B/2h)(\partial B_z/\partial y)\delta\langle y\rangle=12.17$\,MHz, in close agreement with the numerical simulations. The Rabi frequency increases considerably near the spin/valley anti-crossing at $B_y=1.549$\,T, and reaches $f_R=671$\,MHz in the valley qubit regime (independently of $B_y$). This large Rabi frequency results from the in-plane valley dipole induced by alloy disorder. The inter-valley dipole, however, only makes a small contribution to the total displacement $\delta\langle y\rangle$ relevant for the spin qubit regime. Indeed, to first-order in the drive amplitude,
\begin{equation}
\frac{\delta\langle y\rangle}{\delta V_\mathrm{ac}}=e\sum_{n>0}\frac{\langle 0|y|n\rangle\langle n|D_\mathrm{sc}|0\rangle}{E_0-E_n}+\textrm{c.c.}\,,
\end{equation}
where $\ket{n}$ are the spin-up (or spin-down) eigenstates of the qubit (with energies $E_n$) in the absence of spin-orbit coupling. The contribution of the excited valley state $\ket{1}$ is $\delta\langle y\rangle_\mathrm{inter}/\delta V_\mathrm{ac}=0.014$\,nm/mV (over $\delta\langle y\rangle/\delta V_\mathrm{ac}=0.46$\,nm/mV), the series being dominated by the first excited $p_y$ envelope.

The dephasing time follows similar trends. It is approximately constant in the spin qubit regime, where it is expected to be limited by the longitudinal gradients $\partial B_y/\partial z=1.88$\,mT/nm and $\partial B_y/\partial x=-40.73\,\mu$T/nm (as $\partial B_y/\partial y$ is zero along the channel axis in this symmetric layout). Any motion $(\delta\langle x\rangle,\,\delta\langle z\rangle)$ of the dot in these gradients shall indeed result in a drift of Larmor frequency $\delta f_L\approx (g_0\mu_B/h)[(\partial B_y/\partial x)\delta\langle x\rangle+(\partial B_y/\partial z)\delta\langle z\rangle]$ that dephases the qubit. However, the electron hardly moves along the strong confinement axis $z$, and $\partial B_y/\partial x=-40.73\,\mu$T/nm is fairly small (as a matter of fact, the micro-magnet is designed to maximize transverse and minimize dephasing gradients \cite{Philips2022}). Therefore, we find that the above gradient approximation is pretty poor in practice. In particular, the second derivative (curvature) $\partial^2 B_y/\partial y^2$ gives rise to significant corrections to the $\partial f_L/\partial V_i$ of the screening gates that squeeze the dot along $y$ (since the first derivative $\partial B_y/\partial y$ is zero). Nevertheless, the dephasing times of \cref{fig:frTsvsB}, calculated in the exact field of the micro-magnet (no gradient or higher order approximation), remain very long, and matching the experimental spin $T_2^*\approx 5\,\mu$s \cite{Philips2022} calls for large effective noise amplitudes $\delta V_i\sim 100\,\mu$V. The actual nature, density, and dynamics of the fluctuators modulating the electric field in the device remains, however, elusive and beyond the scope of this work. Silicon was isotopically purified in the device of Ref.~\cite{Philips2022} so that we can {\it a priori} rule out strong dephasing by hyperfine interactions \cite{Cvitkovich2024}. We emphasize that the electrical dephasing rate may increase significantly if symmetry is broken because the dot is displaced toward a screening gate, or because the ``upper'' and ``lower'' parts of the micro-magnet are not equivalent, as $\partial B_y/\partial y$ is then finite. For example, $T_2^*$ decreases from $60\,\mu$s to $12\,\mu$s at $B_y=0.5$\,T if the dot is shifted by $\pm 6$\,nm along $y$. The dephasing time drops dramatically ($T_2^*\approx 6$\,ns) in the valley qubit regime, because the valley splitting $\Delta$ is much more dependent on electrical perturbations than the Zeeman splitting.

The spin-phonon relaxation rate continuously increases as $\approx B_y^5$ in the spin qubit regime, because the density of acoustic phonons, the phonon strains, and the phonon wave number at the Larmor frequency all scale as powers of $f_L\propto B_y$. The scaling would be different without micro-magnets, since time-reversal symmetry constraints give rise to an additional $\propto B_y^2$ dependence with intrinsic spin-orbit coupling only (thus a net $\propto B_y^7$ behavior, see Appendix \ref{app:eph} and Ref. \cite{Li2020}). The relaxation rate increases considerably in the valley qubit regime due to the significant inter-valley dipole matrix elements. Strikingly, the shear strain component of the phonons, which couples the $\pm Z$ valleys [\cref{eq:hstrains}], does not make sizable contributions to the relaxation, even in the valley qubit regime. In fact, the matrix elements of $H_\mathrm{shear}$ between $\pm Z$ valley envelopes $\tilde{\varphi}_-(\br)=e^{-ik_0'z}\varphi(\br)$ and $\tilde{\varphi}_+(\br)=e^{ik_0'z}\varphi(\br)$ are, for long wavelength acoustic phonons, proportional to the $2k_0'$ Fourier component of the density $|\varphi(\br)|^2$ \cite{Woods2024}, which is small unless the confinement potential itself is rapidly varying enough. Consequently, the effects of shear strains are more sensible with, e.g., Ge spikes (or in wiggle wells with wave number $q=2k_0'$). Moreover, $\Xi_s$ mostly contributes to the $\Lambda^\mathrm{B}$ terms of Eq.~\eqref{eq3Drelax} for $\Gamma_1$ [see the definitions of $\Lambda^\mathrm{B}_{1l}$ and $\Lambda^\mathrm{B}_{1t}$ in Eqs.~\eqref{eq:lambdaB1} and \eqref{eq:lambdaB2}]. However, these terms scale as $f_L^3$, while the $\Lambda^\mathrm{A}$ terms featuring the hydrostatic and uniaxial deformation potentials $\Xi_d$ and $\Xi_u$ scale as $f_L^5$. The effects of the shear component of the phonons are thus more prominent at small Larmor frequencies.

\begin{figure}
    \centering
    \includegraphics[width=\columnwidth]{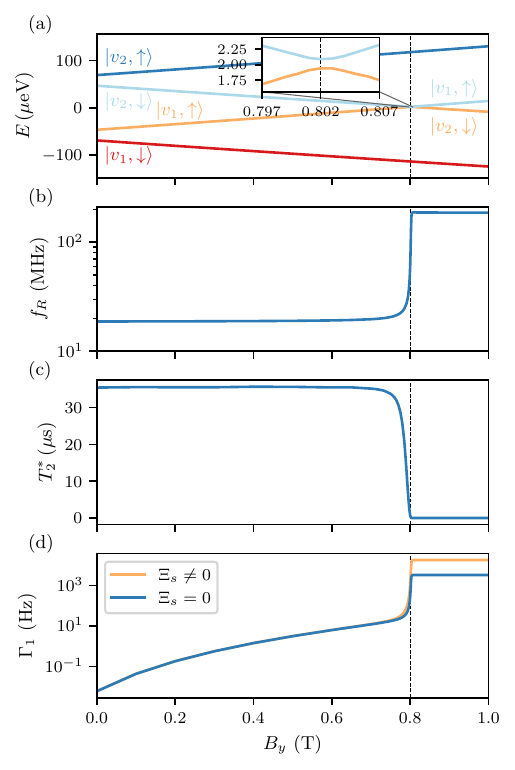}
    \caption{Spin qubit in a quantum well with a Ge spike. (a) Energy levels of the qubit as a function of the external magnetic field $B_y$ along $y$. (b) Rabi frequency $f_R$ as a function of $B_y$ (driving a screening gate with a signal of amplitude $\delta V_\mathrm{ac}=1$\,mV). (c) Dephasing time $T_2^*$ as a function of $B_y$ assuming noise amplitudes $\delta V_i=10\,\mu$V on all gates. (d) Spin-phonon relaxation rate $\Gamma_1$ calculated with and without the shear strain deformation potential $\Xi_s$, at temperature $T=100$\,mK. The vertical dashed line in all panels indicates the critical magnetic field at which the Zeeman energy exceeds the valley splitting.}
    \label{fig:frTsvsB2}
\end{figure}

This is illustrated in \cref{fig:frTsvsB2} for the quantum well with a Ge spike ($\Delta=115\,\mu$eV). The trends are qualitatively similar to \cref{fig:frTsvsB}, yet the Rabi frequencies and dephasing rates in the spin qubit regime are slightly larger in this device because the displacements $\delta\langle y\rangle$ and $\delta\langle z\rangle$ induced by the gates are also stronger with the spike. The effects of the shear component of the phonons are, indeed, now sizable in the valley qubit regime. For Ge spikes, we can compare the performances of the qubit with and without alloy disorder, as the valley splitting remains significant in both cases (here $\Delta=114\,\mu$eV without alloy disorder). While disorder has little impact in the spin qubit regime (except near the spin/valley anti-crossing), it lowers the Rabi frequency, dephasing and relaxation rates by orders of magnitude in the valley qubit regime (in the absence of alloy disorder, $f_R$ is for example smaller in the valley than in the spin qubit regime). Alloy disorder is thus expected to play a major role wherever the valley can interfere with the spin degrees of freedom, such as in shuttling or readout.

\section{Conclusion}

We have implemented an inter-valley potential in the two bands \kp model for the $X$, $Y$ and $Z$ valleys of silicon and SiGe heterostructures. This potential must be defined on a grid whose step is the monolayer thickness $a/4$ along the valley axis, but can be much coarser perpendicular to that axis. Therefore, the two bands \kp model is far more efficient than tight-binding (which effectively meshes the materials at the atomic scale in all directions), and is therefore well suited to the description of large scale structures such as Si/SiGe spin qubit devices. The two bands \kp model also captures the effects of shear strains on the valleys (missing in the effective mass approximation) and the bulk Dresselhaus interaction around the $X$, $Y$ and $Z$ points (when spin is included). We have, moreover, discussed the description of alloy disorder in the two bands \kp model. We have parametrized the model against TB calculations, and analyzed, in particular, the nature of the inter-valley potential in SiGe alloys (and clarified its relations with the band offset). The statistics of the valley splittings and inter-valley dipole matrix elements calculated with the two bands \kp in relevant, disordered SiGe heterostructures are in very good agreement with multi-orbital tight-binding models. This demonstrates that the two bands \kp model captures valley-orbit mixing effects when alloy disorder is playing a leading role. We have applied this model to a realistic Si/SiGe spin qubit and discussed spin and valley physics in this device. In particular, we have derived the expression for spin/charge scattering by phonons, including the (usually neglected) shear strain component. We have also highlighted the importance of alloy disorder in the dynamics of valley qubits and, by extension, of spin qubits where spin-valley mixing is significant. This model opens the way for efficient and accurate simulations of Si/SiGe spin qubits and shuttlers, and, more generally, silicon devices where spin and valley physics are relevant.

\section*{Acknowledgements}

We thank Esteban Rodriguez-Mena for a careful reading of the manuscript. This work was supported by the ``France 2030'' program (PEPR PRESQUILE-ANR-22-PETQ-0002), and by the Horizon Europe Framework Program (grant agreement 101174557 QLSI2). Part of the calculations were performed using computational resources provided by GENCI–IDRIS (Grant 2026-A0190912036).

\appendix

\section{The coefficient $A$ in tight-binding models}
\label{app:TB}

In TB models, the Bloch waves are expanded in a basis set of orthogonal orbitals $\chi_\beta(\br-\vec{R}_{j\alpha})$, where $\beta\in\{s,\,p_x,\,p_y,\,p_z,\,...\}$, $j$ labels the unit cells of the diamond lattice, and $\alpha\in\{1,\,2\}$ labels the two sublattices:
\begin{equation}
e^{\pm ik_0z}u_\pm(\br)\equiv\sum_{j,\alpha,\beta}c_{\alpha\beta}^\pm e^{\pm ik_0z_{j\alpha}}\chi_\beta(\br-\vec{R}_{j\alpha})\,.
\end{equation}
The TB coefficients $c_{\alpha\beta}^\pm$ satisfy the relation $c_{\alpha\beta}^-=(c_{\alpha\beta}^+)^*$ and are normalized so that $\sum_{\alpha,\beta}|c_{\alpha\beta}^\pm|^2=1$. Let us now consider the valley wave functions $\psi_\pm(\br)=\varphi(\br)\sqrt{2\omega_\mathrm{at}}e^{\pm ik_0z}u_\pm(\br)$ drawn from a common envelope $\varphi(\br)$ in a potential $V(\br)$ (the factor $\sqrt{2\omega_\mathrm{at}}$ with $\omega_\mathrm{at}=a^3/8$ the atomic volume ensuring normalization when $\int d^3\br\,|\varphi(\br)|^2=1$). If $V(\br)$ is slowly varying at the atomic scale, then its representation in the TB basis set is approximately diagonal:
\begin{align}
\langle\chi_{\beta'}(\br-\vec{R}_{j'\alpha'})|&V|\chi_\beta(\br-\vec{R}_{j\alpha})\rangle\approx \nonumber \\
&V(\vec{R}_{j\alpha})\delta_{jj'}\delta_{\alpha\alpha'}\delta_{\beta\beta'}
\end{align}
so that:
\begin{align}
J&=\langle\psi_+|V|\psi_-\rangle \nonumber \\
&\approx 2\omega_\mathrm{at}\sum_{j,\alpha,\beta}e^{-2ik_0z_{j\alpha}}V(\vec{R}_{j\alpha})|\varphi(\vec{R}_{j\alpha})|^2(c_{\alpha\beta}^+)^*c_{\alpha\beta}^-\,.
\end{align}
This can be further approximated as a real-space integral:
\begin{equation}
J\approx A\int d^3\br\,e^{-2ik_0 z}\,V(\br)\left|\varphi(\br)\right|^2
\label{eq:JTB}
\end{equation}
with:
\begin{equation}
A=2\sum_{\beta}(c_{1\beta}^+)^*c_{1\beta}^-=2\sum_{\beta}(c_{2\beta}^+)^*c_{2\beta}^-=\sum_{\alpha,\beta}(c_{\alpha\beta}^+)^*c_{\alpha\beta}^-
\label{eq:ATB}
\end{equation}
since sublattices 1 and 2 are equivalent. For the TB model of Ref.~\cite{Niquet2009}, we get $|A|=0.175$, which is smaller than the value $|A|=0.26$ obtained with {\it ab initio} Bloch functions \cite{Saraiva2011}. This partly results from the slightly different approximations made in \cref{eq:Jfinal} and \cref{eq:JTB}. Nevertheless, the value $|A|\approx 0.225$ provides the best agreement between the two bands \kp model and the TB calculations of \cref{sec:assumptions}, which include effects beyond \cref{eq:JTB}, in particular about the form of $\psi_\pm$.

We would like to emphasize that minimal TB models with only one orbital per atom (such as the model of Ref. \cite{Boykin2004}) yield by design $|A|=2|(c_{1s}^+)^*c_{1s}^-|=2|c_{1s}^+|^2=1$. They hence suggest that $V_\mathrm{inter}(\br)=V(\br)$, thus that $V_\mathrm{inter}(\br)$ is the conduction band offset potential $V_\mathrm{CBO}(\br)$ in SiGe alloys. This is, however, not backed by multi-orbital TB models and by {\it ab initio} calculations. Moreover, the substitution of Si by Ge atoms is not accounted for by a diagonal potential in TB (as it also modifies nearest neighbor interactions), and thus can not be described by \cref{eq:JTB,eq:ATB}. We discuss in Appendix \ref{app:Geatoms} the inter-valley potential in a SiGe alloy, and its parametrization on TB models.

\section{$2k_0$ theory in a SiGe alloy}
\label{app:Geatoms}

In section \ref{sec:methodology}, we have derived the expression of the inter-valley matrix element $J$ assuming that the potential $V(\br)$ is slowly varying at the scale of the unit cell. This is obviously not the case for the substitution of a Si by a Ge atom. In this appendix, we derive $J$ for SiGe alloys, and discuss the implications for the two bands \kp model.

We thus assume that $V(\br)=\sum_n \delta V(\br-\br_n)$ is the sum of atomic potentials $\delta V(\br-\br_n)$ describing the substitution of Si by Ge atoms at positions $\{\br_n\}$. We also assume that these potentials are sufficiently short-range so that the envelope functions do not vary significantly on the scale of $\delta V$. Then,
\begin{align}
J&=\langle\psi_+|V|\psi_-\rangle \nonumber \\
&=\sum_n \int d^3\br\,e^{-2ik_0z}\delta V(\br-\br_n)|\varphi(\br)|^2u_+^*(\br)u_-(\br) \nonumber \\
&\approx\omega_\mathrm{at}\sum_n e^{-2ik_0z_n}V^\mathrm{Ge}(\br_n)|\varphi(\br_n)|^2\,,
\label{eq:JSiGe}
\end{align}
where $\omega_\mathrm{at}=a^3/8$ is the atomic volume and:
\begin{align}
V^\mathrm{Ge}&(\br_n) \nonumber \\
&=\frac{1}{\omega_\mathrm{at}}\int d^3\br\,e^{-2ik_0(z-z_n)}\delta V(\br-\br_n)u_+^*(\br)u_-(\br) \nonumber \\
&=\frac{e^{i\frac{4\pi}{a}z_n}}{\omega_\mathrm{at}}\int d^3\br\,e^{2ik_0'(z-z_n)}\delta V(\br-\br_n)\tilde{u}_-^*(\br)\tilde{u}_+(\br)\,.
\label{eq:Vn}
\end{align}
Any translation of the position $\br_n$ of a Ge atom by a lattice vector leaves $V^\mathrm{Ge}(\br_n)$ invariant since the Bloch functions are periodic on the diamond lattice. The integrals $V^\mathrm{Ge}(\br_n)$ can, therefore, only depend on the sublattice where the Si atom is substituted. As the Bloch functions at the $Z$ point satisfy the symmetry relation $\tilde{u}_-^*(\br+\br_s)\tilde{u}_+(\br+\br_s)=-\tilde{u}_-^*(\br)\tilde{u}_+(\br)$ with $\br_s=a(1,\,1,\,1)/4$ the translation from one sublattice to the other, it can further be shown that $V^\mathrm{Ge}(\br_n)\equiv -i V_\mathrm{inter}^\mathrm{Ge}$ is independent on the position of the Ge atoms ($V^\mathrm{Ge}(\br_n)$ being imaginary with the phase conventions used for the Bloch functions of the two bands \kp Hamiltonian).

We thus reach:
\begin{equation}
J\approx\omega_\mathrm{at}\sum_n [-ie^{-i\frac{4\pi}{a}z_n}V_\mathrm{inter}^\mathrm{Ge}][ e^{2ik_0'z_n}|\varphi(\br_n)|^2]\,.
\end{equation}
The discretization of this equation on the finite difference mesh (along the lines of \cref{sec:implementation2kp}) yields \cref{eq:Jfd} and \cref{eq:Vinter} with
\begin{equation}
V_\mathrm{inter}(\br_{ijk})=V_\mathrm{inter}^\mathrm{Ge}\frac{n_{ijk}\omega_\mathrm{at}}{\omega_{ijk}}\,,
\end{equation}
and $n_{ijk}$ the number of Ge atoms in the elementary box $\hat{\omega}_{ijk}$ around $\br_{ijk}$, or equivalently:
\begin{equation}
V_\mathrm{inter}(\br_{ijk})=V_\mathrm{inter}^\mathrm{Ge}Y(\br_{ijk})
\end{equation}
with $Y(\br_{ijk})$ the Ge concentration at $\br=\br_{ijk}$.

$V_\mathrm{inter}^\mathrm{Ge}$ can be estimated from the valley splitting $\Delta$ in a silicon supercell with one single Ge atom ($V_\mathrm{inter}^\mathrm{Ge}=\Delta N/2$ with $N$ the total number of atoms in the supercell). Tight-binding calculations in supercells biaxially strained onto a Si$_{1-X}$Ge$_X$ buffer and relaxed with Keating's valence force field yield $V_\mathrm{inter}^\mathrm{Ge}=511$\,meV for $X=20$\%, $V_\mathrm{inter}^\mathrm{Ge}=514$\,meV for $X=30$\% and $V_\mathrm{inter}^\mathrm{Ge}=517$\,meV for $X=40$\%. $V_\mathrm{inter}^\mathrm{Ge}$ is thus little dependent on the buffer concentration and is indeed very close to the band offset between Si and Ge biaxially strained on that buffer ($\approx 570$\,meV according to \cref{tab:parameters}) \cite{Losert2023}. This correspondence however looks fortuitous and may not hold in other alloys such as SiC or SiGeC, given that the intra-valley band offset has a different ($\propto \int d^3\br\,\delta V(\br)|\tilde{u}_\pm(\br)|^2$) and {\it a priori} unrelated expression.

Moreover, we would like to emphasize that the range of the potential $\delta V(\br)$ is not strictly limited to first nearest neighbors even in the TB model due to the local strains introduced by the Ge atom. Without such local strains (namely, in purely biaxial strains) $V_\mathrm{inter}^\mathrm{Ge}\approx 380$\,meV is much smaller. The fact that the parameter $V_\mathrm{inter}^\mathrm{Ge}=514$\,meV accounts for the effects of these inhomogeneous strains explains why only the average biaxial strains shall be input in the two bands \kp model in disordered alloys. However, the ``short-range'' assumption of \cref{eq:JSiGe} is not fully satisfied in practice due to the local strains. Yet \cref{sec:validation} shows that $V_\mathrm{inter}^\mathrm{Ge}=514$\,meV reproduces the TB valley splittings in a wide range of relevant heterostructures. Nevertheless, the model may break down at large concentrations where the contributions of the Ge atoms may not be additive any more.

\section{Electron-phonon coupling in the four bands bands model}
\label{app:eph}

In this appendix, we give the expressions for the charge/spin-phonon scattering rates in the four bands \kp model (including the shear strain deformation potential $\Xi_s$). We consider a simple ansatz for bulk (3D) acoustic phonons, comprising a single longitudinal branch ($l$) with velocity $v_l$ and two degenerate transverse branches ($t_1$, $t_2$) with velocity $v_t$.

Let
\begin{subequations}
\begin{align}
\langle\br|\zero\rangle&=\sum_{i\in\{+,-\}}\sum_{\sigma\in\{\uparrow,\downarrow\}} a_{i\sigma}(\br)\tilde{u}_{i\sigma}(\br) \\
\langle\br|\one\rangle&=\sum_{i\in\{+,-\}}\sum_{\sigma\in\{\uparrow,\downarrow\}} b_{i\sigma}(\br)\tilde{u}_{i\sigma}(\br)
\end{align}
\end{subequations}
be two states with envelopes $a_{i\sigma}(\br)$ and $b_{i\sigma}(\br)$ on the $\{\tilde{u}_{-\uparrow},\tilde{u}_{-\downarrow},\tilde{u}_{+\uparrow},\tilde{u}_{+\downarrow}\}$ Bloch functions, and energy splitting $E_{01}=\hbar\omega$.

\begin{widetext}
Following Ref.~\cite{Li2020}, the relaxation rate from state $\ket{\one}$ to state $\ket{\zero}$ is:
\begin{equation}
\Gamma_\mathrm{ph}^\mathrm{3D}=\frac{\omega^3}{8\pi^2\hbar\rho}\coth\left(\frac{\hbar\omega}{2k_BT}\right) \sum_{\alpha\in\{l,t_1,t_2\}} \frac{1}{v_\alpha^5} \int_{0}^{\pi}d\theta\sin\theta\int_{0}^{2\pi}d\varphi\,\Big|\langle \zero | e^{iq_\alpha\hat{\vec{q}}(\theta,\varphi)\cdot\br} \Delta H_\alpha(\theta,\varphi)| \one \rangle \Big|^2\\,
\label{eqgamma3D}
\end{equation}
where $\rho$ is the material density, $\hat{\vec{q}}(\theta,\varphi)$ is the unit vector with polar angle $\theta$ and azimuthal angle $\varphi$, $q_\alpha$ is the phonon wave number such that $v_\alpha q_\alpha=\omega$, and $\Delta H_\alpha(\theta,\varphi)$ is the effective Hamiltonian describing the interaction between an electron and a phonon of branch $\alpha\in\{l,t_1,t_2\}$ with wave vector along $\hat{\vec{q}}$. Using \cref{eq:hstrains},
\begin{subequations}
\label{strain2kp}
\begin{align}
\Delta H_l(\theta,\varphi)&=
\begin{bmatrix}
 \Xi_d+\Xi_u \cos^2\theta & 0& \Xi_s \sin^2\theta \sin 2\varphi & 0\\
 0& \Xi_d+\Xi_u \cos^2\theta & 0 & \Xi_s \sin^2\theta \sin 2\varphi\\
 \Xi_s \sin^2\theta \sin 2\varphi & 0 & \Xi_d+\Xi_u \cos^2\theta &0 \\
 0 & \Xi_s \sin^2\theta \sin 2\varphi & 0 & \Xi_d+\Xi_u \cos^2\theta
\end{bmatrix}\,, \\
\Delta H_{t_1}(\theta,\varphi)&=\frac{1}{2}
\begin{bmatrix}
 -\Xi_u \sin2\theta & 0& \Xi_s \sin2\theta \sin 2\varphi & 0\\
 0& -\Xi_u \sin2\theta & 0 & \Xi_s \sin2\theta \sin 2\varphi\\
 \Xi_s \sin2\theta \sin 2\varphi & 0 & -\Xi_u \sin2\theta &0 \\
 0 & \Xi_s \sin2\theta \sin 2\varphi & 0 & -\Xi_u\sin2\theta
\end{bmatrix}\,, \\
\Delta H_{t_2}(\theta,\varphi)&=
\begin{bmatrix}
 0 & 0& \Xi_s \sin\theta \cos 2\varphi & 0\\
 0& 0 & 0 & \Xi_s \sin\theta \cos 2\varphi\\
 \Xi_s \sin\theta \cos 2\varphi & 0 & 0 &0 \\
 0 & \Xi_s \sin\theta \cos 2\varphi & 0 & 0
\end{bmatrix}\,.
\end{align}
\end{subequations}
\end{widetext}
We can complete the integration over $\theta$ and $\varphi$ in Eq.~(\ref{eqgamma3D})  using the dipole approximation (see Appendix A of Ref.~\cite{Li2020}). We end up with
\begin{align}
\Gamma_\mathrm{ph}^\mathrm{3D}&=\frac{\omega^3}{8\pi\hbar\rho}\coth\left(\frac{\hbar\omega}{2k_BT}\right) \nonumber \\
&\times\sum_{\alpha\in\{l,t\}}\left(\frac{\omega^2}{v_\alpha^7}\sum_{n=1}^5 A_n\Lambda^\mathrm{A}_{n\alpha}+\frac{1}{v_\alpha^5} B_1\Lambda^\mathrm{B}_{1\alpha}\right)
\label{eq3Drelax}
\end{align}
where the $\Lambda$'s depend on material parameters and the $A$'s and $B$'s can be expressed as a function of the following moments of the electronic envelopes:
\begin{subequations}
\begin{align}
S_{ij}&=\int d^3\br \left[a^*_{i\uparrow}(\br) b_{j\uparrow}(\br) + a^*_{i\downarrow}(\br) b_{j\downarrow}(\br)\right] \\
R^k_{ij}&=\int d^3\br \left[a^*_{i\uparrow}(\br) b_{j\uparrow}(\br) + a^*_{i\downarrow}(\br) b_{j\downarrow}(\br)\right]r_k \\
T^{kk'}_{ij}&=\int d^3\br \left[a^*_{i\uparrow}(\br) b_{j\uparrow}(\br) + a^*_{i\downarrow}(\br) b_{j\downarrow}(\br)\right]r_k r_{k'}
\end{align}
\label{eqmoments}
\end{subequations}
and:
\begin{equation}
O^{mn}_{ijkl}=R^m_{ij}R^{n*}_{kl}-\frac{1}{2}\left(T^{mn}_{ij}S_{kl}^* + T^{mn*}_{kl}S_{ij}\right)\,.
\label{eqOmijkl}
\end{equation}
Namely,
\begin{subequations}
\begin{align}
A_1&=\sum_{i,j}O^{xx}_{iijj} + O^{yy}_{iijj} \\
A_2&=\sum_{i,j}O^{zz}_{iijj} \\
A_3&=\sum_{i \neq j, k \neq l} O^{xx}_{ijkl} + O^{yy}_{ijkl} \\
A_4&=\sum_{i \neq j, k \neq l} O^{zz}_{ijkl}\\
A_5&=\sum_{i, j \neq k} O^{xy}_{iijk} + O^{xy}_{jkii} + O^{yx}_{iijk} + O^{yx}_{jkii}
\end{align}
\end{subequations}
and
\begin{align}
B_1&=\sum_{i \neq j, k \neq l} S_{ij}S^*_{kl}\,.
\label{eqBs}
\end{align}
The sums over $i,j,k,l$ run over $\{+,-\}$. The $\Lambda^\mathrm{A}_{nl}$ parameters for longitudinal phonons are:
\begin{subequations}
\begin{align}
\Lambda^\mathrm{A}_{1l}&=\frac{4{{\Xi_u}^{2}}}{35}+ \frac{8{\Xi_d\Xi_u}}{15}+\frac{4{\Xi_d}^{2}}{3} \\
\Lambda^\mathrm{A}_{2l}&=\frac{4{\Xi_u}^{2}}{7}+\frac{8\Xi_d\Xi_u}{5}+ \frac{4{\Xi_d}^{2}}{3} \\
\Lambda^\mathrm{A}_{3l}&=\frac{16{{\Xi_s}^{2}}}{35} \\
\Lambda^\mathrm{A}_{4l}&=\frac{16{{\Xi_s}^{2}}}{105} \\
\Lambda^\mathrm{A}_{5l}&=\frac{8{{\Xi_u}{\Xi_s}}}{105}+\frac{8\Xi_d\Xi_s}{15}
\end{align}
\end{subequations}
while the $\Lambda^\mathrm{A}_{nt}$ parameters for transverse phonons are:
\begin{subequations}
\begin{align}
\Lambda^\mathrm{A}_{1t}&=\frac{16{{\Xi_u}^{2}}}{105} \\
\Lambda^\mathrm{A}_{2t}&=\frac{8{{\Xi_u}^{2}}}{35} \\
\Lambda^\mathrm{A}_{3t}&=\frac{64{{\Xi_s}^{2}}}{105} \\
\Lambda^\mathrm{A}_{4t}&=\frac{8{{\Xi_s}^{2}}}{21} \\ 
\Lambda^\mathrm{A}_{5t}&=-\frac{8{{\Xi_u}{\Xi_s}}}{105}\,.
\end{align}
\end{subequations}
Finally, the $\Lambda^\mathrm{B}_{nl}$ and $\Lambda^\mathrm{B}_{nt}$ parameters are:
\begin{align}
\Lambda^\mathrm{B}_{1l}&=\frac{16 {{\Xi_s}^{2}}}{15}
\label{eq:lambdaB1}
\end{align}
and:
\begin{align}
\Lambda^\mathrm{B}_{1t}&=\frac{24{{\Xi_s}^{2}}}{15}\,.
\label{eq:lambdaB2}
\end{align}
We fall back to the expressions of Ref.~\cite{Tahan2014} if $\Xi_s=0$. Note that the prefactors of the $A$ and $B$ terms in Eq.~\eqref{eqgamma3D} scale differently with respect to the level splitting $\hbar\omega$ (as $\omega^5$ and $\omega^3$ respectively). The $S_{ij}$'s, $R_{ij}^k$'s and $T_{ij}^{kk'}$'s also scale as $\omega$ in the spin qubit regime in the absence of micro-magnet (because intrinsic spin-orbit interactions can not couple opposite spins to an electric or strain field if time-reversal symmetry is not broken by a finite magnetic field) \cite{Li2020}. In the main text, we use the parameters for Si in the whole heterostructure: $v_l=9000$\,m/s, $v_t=5400$\,m/s, and $\rho=2329$\,kg/m$^3$ \cite{Li2020}.

\bibliography{references.bib}

\end{document}